\documentclass[bibyear]{aa} 
\usepackage[utf8]{inputenc}
\usepackage[T1]{fontenc}
\usepackage{placeins}
\usepackage[varg]{txfonts}
\usepackage{amsmath,amssymb}
\usepackage{bm}
\usepackage{graphicx}
\usepackage{enumitem}

\usepackage{xcolor}
\usepackage{natbib}
\usepackage{hyperref}
\hypersetup{
  colorlinks = true, 
  urlcolor  = blue, 
  linkcolor = red,  
  citecolor = blue  
}
{\bfseries}{\itshape}
\makeatletter

\newcommand{\dS}{d_{\mathrm{S}}}
\newcommand{\dL}{d_{\mathrm{L}}}
\newcommand{\kS}{k_{\mathrm{S}}}
\newcommand{\kL}{k_{\mathrm{L}}}
\newcommand{\KS}{K_{\mathrm{S}}}
\newcommand{\KL}{K_{\mathrm{L}}}
\newcommand{\LS}{L_{\mathrm{S}}}
\newcommand{\muL}{\mu_{\mathrm{L}}}
\newcommand{\sigL}{\sigma_{\mathrm{L}}}
\newcommand{\ZL}{Z_{\mathrm{L}}}
\newcommand{\ip}[2]{\left\langle #1,\, #2 \right\rangle_{W}}
\newcommand{\tr}{\operatorname{tr}}
\newcommand{\vecop}{\operatorname{vec}}
\newcommand{\wFBET}{\textsc{wFBET}}

\usepackage{orcidlink}
\title{Gram--Schmidt correlation bookkeeping for weighted FBET-type
frameworks: separating short- and long-range correlation scales in
correlated measurements, with an application to radio luminosity
functions}
\titlerunning{Gram--Schmidt correlation bookkeeping}
\authorrunning{Imbri\v{s}ak and Tisani\'c}
\author{Marko Imbri\v{s}ak\thanks{\emph{marko.imbrisak@gmail.com}}\inst{1}\orcidlink{0000-0002-2773-8617}
\and Kre\v{s}imir Tisani\'c\inst{1}\orcidlink{0000-0001-6382-4937}}

\institute{Independent researcher, Zagreb, Croatia}
\date{\today}
\abstract{
Uncertainty quantification for correlated measurements
frequently
requires a covariance model that mixes several qualitatively different
correlation structures: local measurement noise, short-range
(e.g.\ cyclic or seasonal) correlations, and long-range
correlations
associated with trend, drift, or long-memory behavior. Weighted and
generalized least squares can encode such structure only when
the covariance matrix is known or credibly specified in advance, which
in practice it rarely is. We describe a structured
\emph{correlation bookkeeping} construction intended for use inside a
weighted FBET-type (\wFBET{}) uncertainty-quantification
framework: candidate correlation basis functions associated
with distinct lag variables are orthogonalized with a weighted
Gram--Schmidt procedure (under an explicitly declared,
statistically motivated metric) before they are allowed to enter
the covariance or metric construction.
This avoids the double counting inherent in naive additive covariance
decompositions of the form $\Sigma = \Sigma_{\mathrm{short}} +
\Sigma_{\mathrm{long}}$.
We benchmark the construction (in its vectorized form, its
operator-valued kernel form, and the finite-rank basis-function
realization used by the implementation) on the classic
AirPassengers benchmark dataset, and we apply it to the astrophysical
case of radio luminosity functions measured in redshift bins.
No claim of universal optimality is made;
the goal is a transparent and auditable way to organize correlation
scales within weighted information-geometric fitting frameworks.}

\begin{document}

\maketitle
\nolinenumbers

\section{Introduction}
\label{sec:intro}

Practical uncertainty models for correlated measurements are rarely
built from a single correlation mechanism: local measurement or
sampling noise, short-range (cyclic or otherwise) correlations, and
long-range drifts or long-memory behavior typically coexist
\citep{beran1994,cressie1993,rasmussen2006}. When a fitted model is
subsequently used for uncertainty propagation, the attribution of
uncertainty to these different scales matters: a covariance model can
reproduce the total uncertainty correctly while booking variance that
in fact originates from short-range correlations under the
long-range component (or vice versa), and any conclusion drawn from
the individual components then inherits this misattribution.

The standard tools for encoding correlated uncertainties are weighted
and generalized least squares. Both assume
that a covariance matrix
is known or credibly specified. Writing $y \in \mathbb{R}^n$ for the
observations, $f(\theta)$ for a model with parameters $\theta$, and
$\Sigma$ for the residual covariance, the generalized least-squares
objective is
\begin{equation}
\chi^2(\theta) = r(\theta)^{\mathsf{T}} r(\theta),
\qquad
r(\theta) = \Sigma^{-1/2} \bigl( y - f(\theta) \bigr),
\label{eq:wls}
\end{equation}
in which the residuals are standardized by the covariance. Equation
\eqref{eq:wls} is transparent and well understood, and we treat it
throughout as the baseline against which any structured construction
must be compared. The practical difficulty is upstream of
\eqref{eq:wls}: the construction of $\Sigma$ itself is ambiguous
whenever several correlation mechanisms plausibly coexist. It is
tempting to write a sum of per-mechanism covariances, but (as
discussed in detail in Section~\ref{sec:gs}) candidate correlation
basis functions built from different lag variables are generally
\emph{not} orthogonal to one another, and adding them
double-counts whatever structure they share.

This paper develops a structured alternative, intended as a component
of a weighted FBET-type (\wFBET{}) uncertainty framework. FBET, the
Full Bayesian Evaluation Technique \citep{leeb2008}, is an iterative
generalized-least-squares evaluation scheme developed in nuclear data
evaluation \citep[see also][]{schnabel2018,schnabel2021}, in which
prior information and measurement
covariance are combined in a linearized Bayesian update that yields a
full covariance estimate on a chosen grid. The weighting idea
we build on comes from \citet{imbrisak2025}: dataset- or
channel-specific weights, marginalized over prior distributions,
balance heterogeneous multichannel nuclear cross-section data by
reshaping the metric that the fit operates under. The present work extends the same
bookkeeping philosophy in a different direction: instead of weighting
\emph{datasets}, we organize \emph{correlation scales}. Distinct
correlation structures induce distinct effective directions in the
geometry of the uncertainty model, and these directions should be
orthogonalized (under an explicitly chosen, statistically motivated
metric) before amplitudes are assigned to them.

We emphasize at the outset what is \emph{not} claimed. \wFBET{} as
discussed here is not ordinary weighted least squares, and we do not
reduce it to \eqref{eq:wls}; conversely, we do not claim that the
proposed bookkeeping is a new estimation paradigm, that it is optimal
in any decision-theoretic sense, or that the resulting components are
``independent'' in a probabilistic sense, orthogonality holds only
with respect to the chosen weighted inner product, and only after the
orthogonalization step has actually been performed. The contribution
is deliberately modest: a careful, explicit construction that makes
correlation-scale assumptions auditable rather than implicit.

As a concrete testbed we use the AirPassengers dataset of monthly
totals of international airline passengers, popularized by
\citet{box2015}. A monthly series of this kind combines, in one small
dataset, the correlation mechanisms listed above: (i) approximately
local measurement or sampling noise, (ii) seasonal effects that
couple observations occupying nearby positions in the annual cycle
even when they are separated by many years, and (iii) slow drift or
long-memory behavior that couples observations across the entire
observation window \citep{box2015,beran1994,hosking1981}. The
dataset is small, universally available, strongly
seasonal, and dominated by a monotone growth trend, which makes it a
clean setting in which short-range wrapped-seasonal and long-range
chronological correlation structures are both present and visually
interpretable. A second, astrophysical test case (luminosity-function
measurements grouped into channels by redshift bin
\citep{novak2018}) is used in
Section~\ref{sec:prototype}: the redshift bins serve as
independent channels for the Poisson-weighted FBET stage, with no
redshift dependence in the fitted model.

The remainder of the paper follows the structure: background on
weighted and generalized least squares, statistically motivated
weighting, correlation
scales, and the embedding in a \wFBET{}-type framework
(Section~\ref{sec:background}); the selected datasets,
including the AirPassengers coordinate and lag definitions
(Section~\ref{sec:prototype}); the short- and
long-range correlation basis functions (Section~\ref{sec:generators});
the weighted Gram--Schmidt bookkeeping construction and
hyperparameter selection
(Section~\ref{sec:gs}); the results
(Section~\ref{sec:results}); and discussion and conclusions
(Sections~\ref{sec:discussion}--\ref{sec:conclusion}).

\section{Background}
\label{sec:background}

\subsection{Weighted and generalized least squares}
\label{sec:wlsgls}

Let $y \in \mathbb{R}^n$ be observations, $f(\theta) \in \mathbb{R}^n$
a model with parameters $\theta \in \mathbb{R}^{N_p}$, and
$r(\theta) = \Sigma^{-1/2} \bigl( y - f(\theta) \bigr)$ the
standardized residual vector of \eqref{eq:wls}. Under the assumption
that the measurement errors
are Gaussian with known covariance $\Sigma$, maximum likelihood
estimation of $\theta$ reduces to minimizing \eqref{eq:wls}.
Diagonal $\Sigma$ recovers ordinary weighted least squares; a dense
$\Sigma$ gives the generalized form.
The parameter covariance is then estimated, to the usual linearized
approximation, from the inverse of
\begin{equation}
g_{\mu\nu}(\theta)
 = \partial_\mu r(\theta)^{\mathsf{T}}\, \partial_\nu r(\theta),
\label{eq:fim-gls}
\end{equation}
which is the Fisher Information Metric (FIM)
and, via the Cram\'er--Rao bound, a lower bound on estimator
covariance.

The transparency of this baseline is exactly its limitation: all
statistical structure is delegated to $\Sigma$, and \eqref{eq:wls} is
only as credible as the covariance fed into it. When $\Sigma$ must be
assembled from several hypothesized mechanisms (noise, seasonality,
drift) the assembly step is where ambiguity, double counting, and
misattribution enter. The constructions of
Sections~\ref{sec:generators}--\ref{sec:gs} address that assembly
step; they do not modify \eqref{eq:wls} itself.

\subsection{Marginalized channel weights and weighted Fisher
metrics}
\label{sec:wlm}

The channel-weighting technique adopted here was developed by
\citet{imbrisak2025} for
multichannel nuclear cross-section fitting with the Hauser--Feshbach
code CoH$_3$, a setting characterized by ``sloppy'' parameter
directions (FIM eigenvalues spanning many orders of magnitude) and by
strongly imbalanced data: measurement groups (reaction channels,
experimental campaigns) differ widely in size and reported precision.
Measurements are grouped into $N_g$ groups with group-wise residuals
$r^{(k)i}(\theta)$ and statistics $\chi^2_k(\theta) = \sum_i
\bigl(r^{(k)i}(\theta)\bigr)^2$. Each group receives an inverse-variance-like
weight hyperparameter $\alpha_k$, giving the weighted likelihood
\begin{equation}
\mathcal{L}(y \mid \theta, \alpha)
 = \prod_{k=1}^{N_g}
   \left(\frac{\alpha_k}{2\pi}\right)^{n_k/2}
   \exp\!\left(-\tfrac{1}{2}\,\alpha_k\, \chi^2_k(\theta)\right),
\label{eq:wlm-lik}
\end{equation}
after which the $\alpha_k$ are marginalized over a prior $f(\alpha_k)$,
generalizing a weighting technique introduced for joint cosmological
analyses by \citet{hobson2002}. The negative expected Hessian of the
marginalized log-likelihood defines the weighted Fisher Information
Metric,
\begin{equation}
g^{(W)}_{\mu\nu}(\theta)
 = \sum_{k=1}^{N_g} g^{(k)}_{\mu\nu}(\theta)\, w_k(\theta),
\label{eq:wfim}
\end{equation}
where $g^{(k)}_{\mu\nu}$ is the FIM contribution of group $k$ and the
effective weights $w_k(\theta)$ are determined by Laplace-transform
ratios of the prior \citep[Eq.~(15) of][]{imbrisak2025}. In
the original setting this weighted metric balances heterogeneous
groups within an iterative fitting update; here we use only the
weighting construction itself (the marginalized weights $w_k$) not the update scheme built around it. The same weighting machinery has since been
applied to infer unreported measurement uncertainties in
heterogeneous astrophysical datasets \citep{imbrisak2026}.

Two features of this weighting are relevant here. First, the weights are
\emph{statistically motivated}, they arise from marginalization over
explicit priors, not from ad hoc tuning. Second, the weights act
\emph{geometrically}: they reshape the metric on parameter space, and
therefore reshape which directions the uncertainty analysis regards as
stiff or sloppy. The present paper transfers both features from the
dataset index $k$ to a correlation-scale index.

\subsection{Information-geometric interpretation of weights}
\label{sec:infogeo}

In information-geometric terms, the FIM endows parameter space with a
Riemannian metric; model fitting is motion on the resulting manifold,
and (linearized) uncertainty ellipsoids are unit balls of that metric.
Reweighting the likelihood, as in \eqref{eq:wfim}, deforms the metric:
directions informed mainly by down-weighted groups are softened, and
directions informed by up-weighted groups are stiffened. The same
logic applies to correlation structure. A covariance model
$\Sigma(\alpha)$ depending on correlation-amplitude hyperparameters
$\alpha$ changes the standardized residuals entering
\eqref{eq:fim-gls} and
hence changes the geometry seen by the fit. If two correlation
components overlap (in the sense of having a large mutual inner
product under the relevant metric) then their amplitudes are poorly
identified jointly, the induced metric is unstable under
reparametrization between them, and uncertainty attributed to ``the
seasonal part'' versus ``the long-range part'' becomes an artifact of
the parametrization. Orthogonalizing the components under an
explicitly chosen weighted inner product is the direct remedy, and is
the core of Section~\ref{sec:gs}.

\subsection{Short-range and long-range correlations}
\label{sec:shortlong}

We use ``short-range'' and ``long-range'' in the following operational
sense. Short-range correlation acts through a lag variable that is
small for observations regarded as neighbors by some mechanism, here proximity in the \emph{annual cycle}, so that January is close
to both February and December regardless of year. Long-range
correlation acts through the chronological lag over the full
observation window, and is intended to capture slow trend,
demographic or economic growth, persistent drift, and long-memory
behavior of the type studied by \citet{hosking1981} and
\citet{beran1994}. Kernel-based covariance modeling of both kinds is
standard in the Gaussian-process literature
\citep{rasmussen2006}, and the use of variogram- or kernel-generated
covariance families over structured index sets is classical in
spatial statistics \citep{cressie1993}. What is less standard, and
what this paper is about, is the explicit bookkeeping step that
prevents two such kernels, defined over \emph{different} lag
variables on the \emph{same} index set, from double-counting shared
structure when combined.

\subsection{Embedding in a weighted FBET framework}
\label{sec:wfbet}

FBET \citep{leeb2008,schnabel2018,schnabel2021} combines a prior mean $y_0$
with prior covariance $A_0$ and measurements $y_m$ with measurement
covariance $B$ in a linearized Bayesian (iterative generalized-least-squares) update: with a
sensitivity operator $S$, the posterior mean and covariance take the
standard forms
\begin{equation}
\begin{aligned}
y_1 &= y_0 + A_0 S^{\mathsf{T}}
      \bigl( S A_0 S^{\mathsf{T}} + B \bigr)^{-1}
      (y_m - S y_0),\\
A_1 &= A_0 - A_0 S^{\mathsf{T}}
      \bigl( S A_0 S^{\mathsf{T}} + B \bigr)^{-1} S A_0,
\end{aligned}
\label{eq:fbet}
\end{equation}
producing a full covariance estimate on a chosen grid that accounts
for both experimental and model-induced uncertainty, including
reconstruction of unreported correlations. By \wFBET{} we mean, at the
conceptual level used in this paper, an FBET-type evaluation cycle in
which the covariance objects entering \eqref{eq:fbet} (in
particular the prior covariance $A_0$ and/or the measurement
covariance $B$) are not fixed a priori but are assembled from
statistically weighted, structured components, with the weights
playing a role analogous to the marginalized dataset weights of
\citet{imbrisak2025}. We deliberately refrain from deriving update formulas beyond
\eqref{eq:fbet}; the present contribution concerns only the
\emph{assembly} of the structured covariance, not new posterior
algebra.

The sensitivity matrix $S$ is built from Gaussian kernels centered at
the measurement locations, with one bandwidth per grid node; the
bandwidths are fitted by minimizing the $\sigma$-weighted discrepancy
between the data and their smoothed representation, in the spirit of
the Gaussian-process smoothing construction of
\citet{schnabel2018}. The prior pair $(x_0, A_0)$ is formed from
the measurement-normalized kernels and the measurement covariance,
and the update \eqref{eq:fbet} yields the posterior pair
$(x_1, A_1)$ on the grid.
In the weighted variant (the luminosity case), per-channel statistics
$\chi^2_k$ are computed under the current total covariance
$Q + B$, where $Q = S A_0 S^{\mathsf{T}}$ is the prior covariance
propagated to the measurement locations. We emphasize that the
weighted extension presupposes that the measurements decompose into
\emph{several} groups (channels) that can be weighted against one
another; with a single group the weight degenerates to an overall
factor and the update reduces to plain FBET. The AirPassengers series
forms a single measurement group and is therefore analyzed with the
unweighted variant, while the luminosity-function case provides
genuine channel structure (redshift bins) and uses a Poisson
weighting prior. There the channel weights are evaluated from the
Laplace-transform ratios of the weighting prior $f$
\citep[Eq.~(15) of][]{imbrisak2025},
\begin{equation}
\begin{split}
w_k(\theta)
 &= \frac{\mathcal{L}_{s_k}
          \bigl[ \alpha_k^{\,n_k/2+1} f(\alpha_k) \bigr]}
         {\mathcal{L}_{s_k}
          \bigl[ \alpha_k^{\,n_k/2} f(\alpha_k) \bigr]}
 + \frac{\mathcal{L}_{s_k}
          \bigl[ \alpha_k^{\,n_k/2+2} f(\alpha_k) \bigr]}
         {\mathcal{L}_{s_k}
          \bigl[ \alpha_k^{\,n_k/2} f(\alpha_k) \bigr]}\\[4pt]
 &\quad
 - \frac{\mathcal{L}^2_{s_k}
          \bigl[ \alpha_k^{\,n_k/2+1} f(\alpha_k) \bigr]}
         {\mathcal{L}^2_{s_k}
          \bigl[ \alpha_k^{\,n_k/2} f(\alpha_k) \bigr]},
\qquad
s_k = \frac{\chi^2_k(\theta)}{2},
\end{split}
\label{eq:wk}
\end{equation}
where $n_k$ is the number of measurements in channel $k$ and, for a
prior supported on discrete values of $\alpha_k$, the transform is
the sum
$\mathcal{L}_{s}[g(\alpha_k)]
 = \sum_{\alpha_k} e^{-s\,\alpha_k}\, g(\alpha_k)$.
The implementation adopts a Poisson weighting prior with one location
parameter $\mu_k$ per channel,
\begin{equation}
f(\alpha_k)
 = e^{-\mu_k}\, \frac{\mu_k^{\alpha_k}}{\alpha_k!},
\qquad \alpha_k = 0, 1, 2, \dots,
\label{eq:poissonprior}
\end{equation}
and declares two conventions that specialize, or deviate from,
\eqref{eq:wk}: the infinite sum over the discrete $\alpha_k$ is
approximated by truncation at a fixed value, retaining
$\alpha_k = 1, \dots, 4$; and the transform argument is evaluated
as $s_k = \chi^2_k(\theta)/(2N)$ (the per-channel statistic
normalized by the total number of measurements $N$) rather than at
$\chi^2_k(\theta)/2$. Both are conventions of the implementation, and their
influence belongs among the robustness checks of
Section~\ref{sec:diagnostics}; the measurement covariance is
modified per \eqref{eq:Bw} and the update re-run. The Poisson
hyperparameters $\mu_k$ are selected by minimizing the total
FBET $\chi^2$ with the derivative-free search of
Section~\ref{sec:hyperopt}.

The modes of Section~\ref{sec:gs} can enter the \wFBET{} cycle at two
points, corresponding to assemblies (i) and (ii) of
Section~\ref{sec:assembly}.

The FBET update
\eqref{eq:fbet} produces a posterior mean--covariance pair on the
evaluation grid. The basis function families of
Section~\ref{sec:families} are built on the same grid in the
single-index realization, orthogonalized under the declared metric,
and the projection assembly of Section~\ref{sec:assembly}
decomposes the posterior mean and covariance into short-range,
long-range, and cross-scale contributions. The basis function
hyperparameters are selected by the projection-fidelity criterion of
Section~\ref{sec:hyperopt}. In the weighted variant (the
specifically \wFBET{} element) the marginalized channel
weights $w_k$ of \citet{imbrisak2025} modify the measurement covariance entering the update,
\begin{equation}
B \;\longrightarrow\; B^{(w)},
\qquad
B^{(w)}_{ii} = \sigma_i^2 \big/ w_{k(i)},
\label{eq:Bw}
\end{equation}
where $k(i)$ is the channel of measurement $i$ and the $w_k$ are
given by \eqref{eq:wk}; the update \eqref{eq:fbet} is then
re-run with $B^{(w)}$ before the projection bookkeeping is applied.
This follows the workflow that combines FBET smoothing with
marginalized channel weighting, with the correlation
bookkeeping appended as a final attribution step.

\section{Selected datasets}
\label{sec:prototype}

Both analyses are carried out by an implementation whose declared
choices are documented alongside the construction in the following
sections; this section describes the two datasets and their
preprocessing.

\subsection{AirPassengers}

The monthly series \citep{box2015} is
deliberately restricted to a training window narrower than the full
record, covering February 1950 through December 1959 (119 points).
This leaves held-out data on \emph{both} sides of the training
interval (the first thirteen months (January 1949--January 1950)
below it and the final calendar year (1960) above it) so that the
extrapolation of the reconstruction below and above the training
interval can be compared directly against real observations; the
held-out points enter no fit and are used only for this out-of-sample
inspection.
Both coordinates are shifted
to zero minimum ($t \to t - t_{\min}$, $y \to y - y_{\min}$).
Per-point uncertainties are assigned as
$\sigma_i = \sqrt{y_i + s_y^2}$, where $s_y$ is the sample standard
deviation of the series ($y_i$ and $s_y$ are evaluated on the
unshifted passenger counts), a declared heteroscedastic convention
combining a count-like term with a constant floor, not a derived
error model. The FBET evaluation grid consists of $n_g = 11$
equally spaced nodes on the training window. The preprocessed series
is shown as the data points of Fig.~\ref{fig:airpassengers-reproj}.

\paragraph{Seasonal structure.}
\label{sec:airpassengers-bg}
The AirPassengers series \citep{box2015} exhibits (i) a pronounced
annual cycle with a summer peak, (ii) a monotone upward trend across
1949--1960, and (iii) seasonal amplitude that grows with the level of
the series (nonstationary, roughly multiplicative variance). Feature
(i) motivates a short-range basis function on a circular month coordinate;
feature (ii) motivates a long-range basis function on chronological lag;
feature (iii) is a deliberate stress test, since neither basis function is
individually adequate and residual structure remains interpretable.
These properties, together with the dataset's small size and
universal availability, make it a convenient prototyping testbed. We
stress that it is illustrative only: conclusions about the method's
behavior on AirPassengers do not transfer automatically to other
datasets.

\paragraph{Coordinates and lag variables.}
\label{sec:airpassengers}
Let the observations be indexed by $i = 1, \dots, n$ with $n = 144$.
Two equivalent time coordinates are used:
\begin{equation}
\begin{aligned}
t_i &\in \{0, 1, \dots, 143\}
&& \text{(integer month index)},\\
x_i &= \frac{t_i}{12}
&& \text{(normalized years)}.
\end{aligned}
\label{eq:coords}
\end{equation}
The seasonal phase is the fractional part of the normalized time,
\begin{equation}
\phi_i = x_i \bmod 1 \in [0, 1),
\label{eq:phase}
\end{equation}
equivalently described by the month index $m_i = t_i \bmod 12$.

Two lag variables are then defined on pairs $(i, j)$. They are
deliberately \emph{not} collapsed into a single Euclidean distance;
keeping them distinct is the entire point of the construction.

\paragraph{Long-range chronological lag.} In normalized years,
\begin{equation}
\dL(i,j) = \lvert x_i - x_j \rvert,
\label{eq:dL}
\end{equation}
or, if integer months are used, $\dL(i,j) = \lvert t_i - t_j \rvert$.
This lag lives on the half-line $[0, \infty)$ (in practice
$[0, (n-1)/12]$ years) and grows without bound as observations
separate chronologically.

\paragraph{Short-range wrapped seasonal lag.} For any two phases
$\phi, \phi' \in [0, 1)$, define the wrapped distance
\begin{equation}
\dS(\phi, \phi')
 = \min\bigl( \lvert \phi - \phi' \rvert,\;
              1 - \lvert \phi - \phi' \rvert \bigr),
\qquad
\dS(i,j) \equiv \dS(\phi_i, \phi_j),
\label{eq:dS}
\end{equation}
with $\phi_i$ the phase \eqref{eq:phase}; in integer months with
$m_i = t_i \bmod 12$,
\begin{equation}
\dS(i,j)
 = \min\bigl( \lvert m_i - m_j \rvert,\;
              12 - \lvert m_i - m_j \rvert \bigr).
\label{eq:dS-int}
\end{equation}
The wrapped distance \eqref{eq:dS} encodes the circular topology of
the seasonal coordinate: January is close to February \emph{and} to
December, and $\dS$ is bounded by half a period ($1/2$ in phase units,
$6$ in months). Two observations twelve months apart have
$\dS(i,j) = 0$ but $\dL(i,j) = 1$ year; two adjacent observations in
June and July have small values of both. The pair
$\bigl(\dS(i,j), \dL(i,j)\bigr)$ therefore separates ``same place in
the cycle'' from ``close in absolute time,'' which an ordinary
Euclidean distance on $t_i$ cannot do.

The dataset is a clean testbed precisely because both lag variables
are strongly active: the seasonal cycle populates the short-range
structure at all chronological separations, the growth trend populates
the long-range structure, and the multiplicative variance growth
leaves interpretable residual structure under any homoscedastic
model.

\subsection{Luminosity functions}

The second test case is an
astrophysical luminosity-function dataset with measurements
$(x_i, y_i, \sigma_i)$ grouped into channels by redshift bin.
The data are the radio luminosity-function measurements of the
VLA-COSMOS 3\,GHz Large Project \citep{novak2018}: the abscissa is
the rest-frame 1.4\,GHz luminosity,
$\log L_{1.4\,\mathrm{GHz}}\,[\mathrm{W\,Hz^{-1}}]$, the measurements
are grouped into redshift bins following the published binning,
and the fitted ordinate is the luminosity function itself,
$\log\Phi$ (in units of $\mathrm{Mpc^{-3}\,dex^{-1}}$). The
same preprocessing (shift to zero minimum) is applied; the
evaluation grid consists of 15 equally spaced nodes of which one,
falling in an empty interval of the shifted luminosity coordinate, is
removed, leaving $n_g = 14$. The preprocessed measurements are shown
as the data points of Fig.~\ref{fig:lf-reproj}, whose ordinate
carries this label. This dataset has no cyclic coordinate,
so the short-range basis functions use the unwrapped distance
(Section~\ref{sec:families}); its role in the study is to exercise
the interaction between marginalized channel weighting \eqref{eq:Bw}
and the correlation bookkeeping on data with genuine channel
structure.
We emphasize a deliberate methodological simplification: the
redshift bins enter the analysis \emph{only} as independent weighting
channels for the weighted update \eqref{eq:Bw}, effectively as
distinct datasets sharing one abscissa. The fitted model carries no
redshift parameter: a single curve on the (shifted) luminosity axis
is fit to all channels simultaneously, and the well-established
functional dependence of the luminosity function on redshift
(luminosity and density evolution; \citealt{novak2018}) is
intentionally ignored. This is a convention of the demonstration, not
an astrophysical statement: the test targets the interaction of
channel weighting with the correlation bookkeeping, and modeling the
redshift evolution would only obscure that target.

\section{Short-range and long-range correlation basis functions}
\label{sec:generators}

We next define one candidate basis function per correlation scale. We call
them \emph{basis functions} (equivalently, dictionary
elements)
rather than covariance components: they are building blocks for a
correlation model, and they are \emph{not} automatically independent,
orthogonal, or even jointly admissible components. Their promotion to
usable components happens only in Section~\ref{sec:gs}.

\subsection{Short-range wrapped seasonal basis function}

For a seasonal correlation length $\LS > 0$ (in the units of $\dS$),
define
\begin{equation}
\kS(i, j;\, \LS)
 = \LS\, \exp\!\left( - \frac{\dS(i,j)}{\LS} \right).
\label{eq:kS}
\end{equation}
The exponential decay in the \emph{wrapped} lag makes \eqref{eq:kS} a
circular analogue of the familiar Ornstein--Uhlenbeck (exponential)
kernel \citep{rasmussen2006}; because $\dS \le 1/2$ (phase units), the
basis function never fully decays and couples all pairs of months to some
degree, most strongly those adjacent on the annual circle. The
prefactor $\LS$ fixes a reference amplitude scale; since the
Gram--Schmidt step of Section~\ref{sec:gs} renormalizes every
basis function under the working inner product, this prefactor affects
bookkeeping conventions but not the final orthonormalized modes.

\subsection{Long-range chronological basis function}

For the long-range scale we use a lag-domain profile that vanishes at
zero lag, rises as $\sqrt{d}$, is localized around a
characteristic separation $\muL$ with width $\sigL$, and is
normalized directly on $d \ge 0$:
\begin{equation}
\begin{aligned}
\kL(i, j;\, \muL, \sigL)
 &= \frac{\sqrt{\dL(i,j)}\;
    \mathcal{N}\!\bigl(\dL(i,j);\, \muL, \sigL^2\bigr)}
        {\ZL(\muL, \sigL)},\\
\ZL(\muL, \sigL)
 &= \left[ \int_0^\infty u\; \mathcal{N}(u;\, \muL, \sigL^2)\,
   \mathrm{d}u \right]^{1/2},
\end{aligned}
\label{eq:kL}
\end{equation}
where $\mathcal{N}(d;\, \muL, \sigL^2) = (2\pi\sigL^2)^{-1/2}
\exp\bigl[-(d-\muL)^2/(2\sigL^2)\bigr]$ is the Gaussian density.
The basis function is thus the product of a $\sqrt{d}$ envelope and a
Gaussian window, with the overall scale fixed by the normalization on
$[0, \infty)$. The $\sqrt{d}$ factor suppresses the
basis function at small chronological lag (where the short-range
basis function is intended to act) and the Gaussian factor localizes the
long-range coupling around separations of order $\muL$. The integral
defining $\ZL^2$ is available in closed form in terms of the Gaussian
density and cumulative distribution function evaluated at
$-\muL/\sigL$; for $\muL \gg \sigL$ it approaches $\muL$.

The $\sqrt{d}$ envelope is not an aesthetic choice, and a pure
Gaussian bump in the lag would not serve the same purpose. First, the
envelope enforces $\kL \to 0$ as $d \to 0$, so the long-range basis
function carries no weight at small chronological lag, where the
short-range basis function is designed to act; the two scales are
thus partially separated already at the level of the dictionary,
rather than only by the subsequent orthogonalization. A Gaussian
factor alone would instead retain a finite correlation plateau at
zero lag and compete directly with the short-range basis function
there. Second, root-lag growth is the classical signature of
random-walk-like drift: for an intrinsically stationary process with
a linear variogram, $\operatorname{Var}[X(t{+}d)-X(t)] \propto d$, so
the root-mean-square increment grows as $\sqrt{d}$
\citep{cressie1993}, as realized by the Wiener-process covariance
familiar from Gaussian-process regression \citep{rasmussen2006}. The
$\sqrt{d}$ growth also corresponds to the boundary case $H = 1/2$ of
the long-memory family, in which increment amplitudes scale as
$d^{H}$ with Hurst exponent $H$ \citep{hosking1981,beran1994};
steeper or shallower envelopes would encode persistent or
antipersistent drift, and the Gaussian factor truncates the growth by
localizing the coupling around separations of order $\muL$.

The division by $\ZL$ in \eqref{eq:kL} is a \emph{scale convention}:
it fixes a reference normalization of the long-range basis function across
hyperparameter values $(\muL, \sigL)$, so that amplitude
hyperparameters attached to $\kL$ remain comparable as $(\muL,
\sigL)$ vary. It is not a claim of unit norm under the statistical
inner product used later; exact unit norm under that inner product is
imposed by the Gram--Schmidt normalization in Section~\ref{sec:gs},
which supersedes any lag-domain convention.

Both $\kS$ and $\kL$ are functions on the same index-pair set
$\{(i,j)\}$, evaluated through different lag variables. Nothing in
their definitions makes them orthogonal, uncorrelated, or
complementary. On AirPassengers they demonstrably overlap: for
example, pairs of observations one year apart have small $\dS$
(indeed $\dS = 0$) \emph{and} moderate $\dL$, so both basis functions
assign such pairs nonzero correlation. Treating $\kS$ and $\kL$ as
if they were independent covariance components without further
processing double-counts exactly this shared structure.

\subsection{Families of basis functions and single-index realizations}
\label{sec:families}

Two generalizations of the constructions above are used by the
implementation documented in Section~\ref{sec:prototype}
and deserve explicit statement.

\paragraph{Families of basis functions.} Nothing restricts the dictionary
to one basis function per scale. One may take a family of
$N_{\mathrm{S}}$ short-range basis functions with distinct hyperparameters
$\{(m_a, \LS^{(a)})\}_{a=1}^{N_{\mathrm{S}}}$, where $m_a$ is a
reference phase on the seasonal circle, together with
$N_{\mathrm{L}}$ long-range basis functions with hyperparameters
$\{(\muL^{(b)}, \sigL^{(b)})\}_{b=1}^{N_{\mathrm{L}}}$, giving a
dictionary of $N_{\mathrm{S}} + N_{\mathrm{L}}$ elements. The
bookkeeping problem of Section~\ref{sec:gs} then involves overlaps
\emph{within} each family as well as \emph{between} the scales; the
orthogonalization treats both uniformly, and the scale labels survive
as a partition of the mode index used for block bookkeeping in
Section~\ref{sec:assembly}.

\paragraph{Two-index versus single-index realizations.} The
basis functions admit two distinct realizations, which induce different
kernel classes and should not be conflated.
\begin{itemize}[itemsep=2pt]
\item \emph{Two-index (lag-kernel) realization.} The basis function is
evaluated on pairwise lags, $[\KS]_{ij} = \kS(\dS(i,j);\, \LS)$ and
$[\KL]_{ij} = \kL(\dL(i,j);\, \muL, \sigL)$, as written in
\eqref{eq:kS}--\eqref{eq:kL}. The induced correlation structures are
stationary in the respective lag variables.
(The two-index realization is exercised by the operator-valued
check of Sect.~\ref{sec:versionB}.)
\item \emph{Single-index (basis-function) realization.} One argument
of the two-point structure is anchored, and each basis function becomes a
function of a single evaluation coordinate $\xi$, an observation
time or a grid node, expressed in normalized years, entering the
seasonal family through its phase $\xi \bmod 1$. For the short family
the anchor is the learned reference phase $m_a$,
\begin{equation}
\varphi^{(a)}_{\mathrm{S}}(\xi_i)
 = \kS\bigl(\dS(\xi_i \bmod 1,\, m_a);\, \LS^{(a)}\bigr),
\label{eq:single-short}
\end{equation}
a seasonal bump on the annual circle, with $\dS(\cdot,\cdot)$ the
wrapped phase distance of \eqref{eq:dS}; for the long family the
anchor is the series origin,
\begin{equation}
\varphi^{(b)}_{\mathrm{L}}(\xi_i)
 = \kL\bigl(\xi_i;\, \muL^{(b)}, \sigL^{(b)}\bigr),
\label{eq:single-long}
\end{equation}
the lag profile \eqref{eq:kL} read as a function of elapsed time from
the start of the record. Each basis function is then a vector over
observation or grid points, and two-index correlation structure
arises only later, through outer products of the orthogonalized modes
(Section~\ref{sec:assembly}); the induced kernels are of finite rank
(degenerate) rather than stationary.
\end{itemize}
The two realizations are both legitimate but are \emph{not}
equivalent: finite-rank expansions can represent only structure lying
in the span of the basis function family, while stationary lag kernels
constrain the correlation to depend on the lag alone. The implementation
implementation uses the single-index realization; the operator
formulation of Section~\ref{sec:versionB} covers the two-index case.

The wrapped lag $\dS$ applies when the short-range coordinate is
genuinely circular, as for calendar months. For datasets whose
abscissa is not cyclic (such as the luminosity-function test case
of Section~\ref{sec:prototype}) the short-range basis function is used
with the ordinary unwrapped distance
$\lvert \xi - m_a \rvert$ in place of $\dS$. All other elements of
the construction are unchanged.

\section{Weighted Gram--Schmidt correlation bookkeeping}
\label{sec:gs}

This section presents the main technique of the paper: a
searched family of correlation basis
functions is orthonormalized by weighted Gram--Schmidt under an
explicitly declared metric (Sect.~\ref{sec:versionA}), the FBET
posterior pair is projected onto the resulting modes and decomposed
into per-scale and cross-scale blocks
(Sect.~\ref{sec:assembly}), and the basis-function hyperparameters
are selected by a projection-fidelity search
(Sect.~\ref{sec:hyperopt}). Alternative realizations of the same
idea (the operator-valued kernel basis and two amplitude
assemblies) are collected in Sect.~\ref{sec:alternatives} for
comparison.

\subsection{The weighted inner product}

Let $W$ be a symmetric positive-definite weighting matrix on the
relevant space. Statistically motivated choices include: measurement
precision (inverse reported variances, possibly per-point);
an iteratively refined \wFBET{} weight from a previous pass of the
evaluation cycle; or a metric estimate derived from a current
linearization, in the spirit of the weighted-metric construction
\eqref{eq:wfim}. The construction below is stated for a generic
admissible $W$; the choice of $W$ is an explicit modeling assumption
and must be reported alongside any results. The identity $W = I$ is
itself an admissible declared baseline (under it the weighted
procedures below reduce to ordinary Gram--Schmidt, implementable as a
QR factorization) and it is the baseline used by the
implementation of Section~\ref{sec:prototype}. Reporting $W = I$ as
the working metric is legitimate; leaving the metric unstated is not.

\subsection{Vectorized basis and orthonormalized modes}
\label{sec:versionA}

Gram--Schmidt output depends on the processing order. We fix the
convention that the short-range basis function is normalized first and the
long-range basis function is orthogonalized against it, so that the
long-range mode is, by construction, the part of the long-range
basis function not already expressible by the seasonal one. The reverse
convention is equally legitimate and yields different modes; the
choice is a declared bookkeeping convention, and results should be
checked for robustness under swapping it. For any fixed basis function set
the spanned subspace is order-independent even though the individual
modes are not; in a QR implementation the ordering is the column
ordering of the stacked basis function matrix, and with families
(Section~\ref{sec:families}) the convention is short-family-first.

Collect the basis function values over the evaluation
samples (in the single-index realization
\eqref{eq:single-short}--\eqref{eq:single-long}, the observation or
grid points) into vectors (written $h$ to avoid collision with the
seasonal phase $\phi_i$ of \eqref{eq:phase}):
\begin{equation}
h_{\mathrm{S}} = \vecop\bigl( \kS(i,j;\, \LS) \bigr),
\qquad
h_{\mathrm{L}} = \vecop\bigl( \kL(i,j;\, \muL, \sigL) \bigr).
\label{eq:vecbasis}
\end{equation}
On this vector space define the weighted inner product
\begin{equation}
\ip{h_a}{h_b} = h_a^{\mathsf{T}} W\, h_b,
\label{eq:ipvec}
\end{equation}
with $W$ positive definite on the pair space (for instance, built from
products of per-point precisions, or from a \wFBET{} iterate).
Weighted Gram--Schmidt then proceeds in the declared order
(short-range first, the declared ordering convention):
\begin{align}
\psi_1 &= \frac{h_{\mathrm{S}}}
              {\sqrt{\ip{h_{\mathrm{S}}}{h_{\mathrm{S}}}}},
\label{eq:gs1}\\[4pt]
\psi_2^{\mathrm{raw}}
 &= h_{\mathrm{L}} - \ip{\psi_1}{h_{\mathrm{L}}}\, \psi_1,
\qquad
\psi_2 = \frac{\psi_2^{\mathrm{raw}}}
             {\sqrt{\ip{\psi_2^{\mathrm{raw}}}{\psi_2^{\mathrm{raw}}}}}.
\label{eq:gs2}
\end{align}
By construction the resulting short/long correlation basis is
orthonormal under the chosen metric,
\begin{equation}
\ip{\psi_a}{\psi_b} = \delta_{ab},
\qquad a, b \in \{1, 2\},
\label{eq:orthonormal}
\end{equation}
and only \emph{after} \eqref{eq:orthonormal} holds may the modes be
described as orthogonal components, always with the qualifier
``under $\ip{\cdot}{\cdot}$.''

For a family of $N = N_{\mathrm{S}} + N_{\mathrm{L}}$ basis functions
$\{h_1, \dots, h_N\}$ (Section~\ref{sec:families}; short family
first), the procedure iterates in the declared order: each $h_k$
is orthogonalized against $\psi_1, \dots, \psi_{k-1}$ and normalized.
Under the baseline metric $W = I$ this is ordinary Gram--Schmidt and
is implemented most stably as an economy QR factorization of the
stacked basis function matrix, whose orthonormal factor reproduces the
Gram--Schmidt modes up to signs. For general $W$ the implementation
uses the weighted recursion above (with a second orthogonalization
pass restoring orthogonality for nearly degenerate dictionaries); the
diagnostic suite exercises it with the full posterior-precision
metric $A_1^{+}$ in the metric-robustness check. An equivalent alternative route (a QR
factorization of $W^{1/2}\Phi$ followed by the back-map
$\psi = W^{-1/2} q$) yields the same modes and is numerically
preferable for strongly ill-conditioned dictionaries, but is not
required by the implementation. A pre-normalization of
each basis function vector (e.g., to unit Euclidean norm) before
orthogonalization is a harmless convention: it affects conditioning
of the factorization, not the resulting modes.

\subsection{Projection bookkeeping of the posterior pair}
\label{sec:assembly}

Only after orthogonalization may structured correlation objects be
written in terms of the modes. The main technique uses the
projection form below; two alternative amplitude assemblies, which
answer different questions and have different safety properties, are
collected in Sect.~\ref{sec:alternatives} and exercised only in the
Discussion. In
what follows, stack the orthonormalized modes as the rows of
$\Psi \in \mathbb{R}^{N \times n}$, with
$\Psi W \Psi^{\mathsf{T}} = I_N$ by \eqref{eq:orthonormal}, and let
the mode index carry the scale partition
$\{1, \dots, N\} = \mathcal{S} \cup \mathcal{L}$ inherited from the
basis function families.

Suppose a mean--covariance pair $(x, A)$ on the
index set is already available (in the \wFBET{} setting, the
posterior pair produced by the FBET update \eqref{eq:fbet}. Define
mode coefficients and the mode-space covariance
\begin{equation}
c = \Psi W x \in \mathbb{R}^{N},
\qquad
A_{\mathrm{proj}} = \Psi W A\, W \Psi^{\mathsf{T}}
 \in \mathbb{R}^{N \times N},
\label{eq:proj}
\end{equation}
and the reconstructions
\begin{equation}
\begin{aligned}
\widehat{x} &= \Psi^{\mathsf{T}} c,\\
\widehat{C} &= \Psi^{\mathsf{T}} A_{\mathrm{proj}}\, \Psi,\\
\widehat{C}_{ij}
 &= \sum_{a,b=1}^{N} [A_{\mathrm{proj}}]_{ab}\,
   \psi_a(\xi_i)\, \psi_b(\xi_j).
\end{aligned}
\label{eq:reconstruction}
\end{equation}
Since $\widehat{C} = M A M^{\mathsf{T}}$ with
$M = \Psi^{\mathsf{T}} \Psi W$, positive semidefiniteness of
$\widehat{C}$ follows from that of $A$ by congruence; no eigenvalue
check is required. The scale partition gives the block bookkeeping
\begin{equation}
\begin{aligned}
\widehat{C}
 &= \widehat{C}_{\mathcal{S}\mathcal{S}}
 + \widehat{C}_{\mathcal{L}\mathcal{L}}
 + \widehat{C}_{\mathcal{S}\mathcal{L}}
 + \widehat{C}_{\mathcal{L}\mathcal{S}},\\
\widehat{C}_{\mathcal{A}\mathcal{B}}
 &= \sum_{a \in \mathcal{A}} \sum_{b \in \mathcal{B}}
   [A_{\mathrm{proj}}]_{ab}\, \psi_a^{\vphantom{\mathsf{T}}}
   \psi_b^{\mathsf{T}},
\end{aligned}
\label{eq:blocks}
\end{equation}
in which the cross-scale blocks are \emph{retained and reported}, not
assumed away: orthonormality of the modes does not make
$A_{\mathrm{proj}}$ diagonal) that would require the modes to be
eigendirections of $A$ under $W$, so a nonzero
$\widehat{C}_{\mathcal{S}\mathcal{L}}$ is a finding, not an error.
This projection form is the one used by the implementation
of Section~\ref{sec:prototype}. Its per-scale attribution is exact
\emph{within the span} of the basis function family; structure orthogonal
to that span is simply not represented, and the reconstruction
residuals $x - \widehat{x}$ and $A - \widehat{C}$ must be reported as
the price of the compression.

\subsection{Hyperparameter selection}
\label{sec:hyperopt}

The basis function hyperparameters
$\theta = \{m_a, \LS^{(a)}\}_{a} \cup
\{\muL^{(b)}, \sigL^{(b)}\}_{b}$ are selected by minimizing a
projection-fidelity cost measuring how well the mode basis reproduces
the FBET posterior:
\begin{equation}
J(\theta)
 = \bigl\lVert \widehat{x}(\theta) - x_1 \bigr\rVert_2^2
 + \bigl\lVert \widehat{C}(\theta) - A_1 \bigr\rVert_F^2,
\label{eq:projcost}
\end{equation}
with $\widehat{x}, \widehat{C}$ as in \eqref{eq:reconstruction}. (A
variant weighting the mean residual by the pseudo-inverse of the
posterior covariance is also implemented; the choice of cost is a
declared convention.) The minimization uses a derivative-free search
on per-parameter grids: finite-difference gradient and Hessian
estimates are assembled from a stencil of cost evaluations executed
in parallel, a proposal step is taken along the eigenvector
associated with the smallest Hessian eigenvalue with an adaptively
scaled step length (a geometric heuristic in the spirit of the
manifold-based scaling strategy of \citet{imbrisak2025}, applied to
astrophysical uncertainty inference by \citet{imbrisak2026}) and the
step is accepted or rejected on cost decrease. A hyperparameter
covariance estimate is read off from the pseudo-inverse of the final
Hessian. The obvious caveats apply: the optimizer is a heuristic, the
cost surface \eqref{eq:projcost} can be multimodal, no
global-optimality claim is made, and sensitivity of the selected
$\theta$ to the search grids must be reported.

The search grids themselves are declared conventions and
deserve explicit statement. Each hyperparameter is restricted to a
one-dimensional grid of 50 equally spaced points: the seasonal
correlation lengths to $\LS \in [0.1, 1]$, the long-range widths to
$\sigL \in [t_{\max}/n_m,\, t_{\max}]$, and (importantly) the
reference phases $m_a$ and separations $\muL^{(b)}$ to a
\emph{partition} of their ranges, the $a$-th (resp.\ $b$-th) family
member being confined to the $a$-th subinterval of $[0,1)$ (resp.\ of
$[0, t_{\max}]$). The partition is an identifiability device: it
breaks the permutation degeneracy of the family members and prevents
two anchors from collapsing onto the same location, at the price of
forbidding configurations that straddle a subinterval boundary.
Proposal steps of the optimizer snap to the nearest grid point and
the ranges clamp at their boundaries, so the selected $\theta$ can
depend on the discretization itself. The implementation therefore
includes a predeclared grid-sensitivity protocol: the search is
rerun with doubled and halved grid density, with widened $\LS$ and
$\sigL$ ranges, and with unpartitioned anchor grids, reporting the
selected $\theta$ (including boundary hits), the relative Frobenius
change of the per-scale blocks against the baseline grids, and the
ratio of the $\theta$ shifts to the Hessian-based uncertainties, if that ratio materially exceeds unity, the reported hyperparameter
uncertainties are understated.
The results of this protocol are reported in
Sect.~\ref{sec:diagresults} (Fig.~\ref{fig:grid-sensitivity}).

\subsection{Alternative realizations and assemblies}
\label{sec:alternatives}

The constructions collected here are \emph{not} part of the main
pipeline: they are alternative realizations of the same bookkeeping
idea, implemented for comparison and exercised in the Discussion
(Sect.~\ref{sec:discussion}).

\paragraph{Operator-valued kernel basis (Version B).}
\label{sec:versionB}
Alternatively, work directly with the $n \times n$ two-index kernel
matrices $\KS$ and $\KL$ of Section~\ref{sec:families} and
define an operator inner product,
\begin{equation}
\ip{K_a}{K_b} = \tr\bigl( W K_a\, W K_b \bigr),
\label{eq:ipop}
\end{equation}
or, if a manifestly symmetric form is required,
\begin{equation}
\ip{K_a}{K_b}
 = \tr\bigl( W^{1/2} K_a\, W K_b\, W^{1/2} \bigr).
\label{eq:ipop-sym}
\end{equation}
For symmetric $K_a, K_b$ and symmetric positive-definite $W$ the two
forms agree by cyclicity of the trace; \eqref{eq:ipop-sym} makes
positivity of the induced norm explicit, since
$\ip{K}{K} = \lVert W^{1/2} K W^{1/2} \rVert_F^2 \ge 0$ with equality
iff $K = 0$. Gram--Schmidt in operator space reads
\begin{align}
P_1 &= \frac{\KS}{\sqrt{\ip{\KS}{\KS}}},
\label{eq:gsop1}\\[4pt]
P_2^{\mathrm{raw}}
 &= \KL - \ip{P_1}{\KL}\, P_1,
\qquad
P_2 = \frac{P_2^{\mathrm{raw}}}
           {\sqrt{\ip{P_2^{\mathrm{raw}}}{P_2^{\mathrm{raw}}}}},
\label{eq:gsop2}
\end{align}
yielding operator modes $P_1 \equiv P_{\mathrm{S}}$ and
$P_2 \equiv P_{\mathrm{L}}$ with
$\ip{P_a}{P_b} = \delta_{ab}$.
The operator-valued formulation is implemented as a numerical
check: the two-index kernel
matrices are built on the evaluation grid, orthonormalized under the
operator inner product \eqref{eq:ipop}, and used for an
operator-space projection of the posterior covariance; the per-mode
eigenvalue spectra quantify the indefiniteness of the orthogonalized
operator modes that motivates the eigenvalue caveat of assembly (iii)
in Sect.~\ref{sec:assembly}
(Sect.~\ref{sec:diagresults}, Fig.~\ref{fig:operator-gs-winv}).

The two formulations serve different purposes, and the
trade-off deserves explicit statement. The vectorized construction of
Sect.~\ref{sec:versionA} works
with the basis functions as vectors on the evaluation grid: it is
cheap (inner products cost $O(n)$), its modes are directly plottable
functions, and the projection bookkeeping of
Sect.~\ref{sec:assembly} retains the full $N \times N$ mode-space
covariance, so the compression is exact within the span. Its price is
that all two-index structure is finite-rank, built from outer
products of the modes, and tied to the anchoring conventions of the
single-index realization (the reference phases $m_a$ and the series
origin). The operator-valued realization instead orthogonalizes the two-index kernels
themselves: it requires no anchors (the short-range kernel enters
through $\LS$ alone) each dictionary element carries a full
stationary correlation structure in its own lag variable rather than
a rank-one product, and orthogonality is imposed directly between
candidate covariance operators, which is the natural bookkeeping when
the deliverable is a covariance parametrization such as
\eqref{eq:C-safe} or \eqref{eq:C-direct}. Its costs are equally
concrete: operator inner products require $O(n^{3})$ matrix products
and $O(n^{2})$ storage per element; the modes are matrices rather
than curves and are harder to inspect; classical Gram--Schmidt loses
orthogonality numerically when the raw kernels are nearly degenerate
(a second orthogonalization pass restores it); the anchor-free
short-range kernels collapse a family with distinct reference phases
onto the correlation lengths alone, which can make within-family
degeneracy more severe; and, most importantly, the orthogonalized
modes are in general indefinite (only the first mode, a rescaled
admissible kernel, is guaranteed positive semidefinite) which is
precisely why the direct amplitude form \eqref{eq:C-direct} demands
an explicit eigenvalue check.

\paragraph{(ii) Squared amplitude form.} When per-scale amplitudes
are free hyperparameters rather than projections of an existing
covariance, the safe assembly is
\begin{equation}
C(\alpha_{\mathrm{S}}, \alpha_{\mathrm{L}})
 = \alpha_{\mathrm{S}}\, P_{\mathrm{S}} P_{\mathrm{S}}^{\mathsf{T}}
 + \alpha_{\mathrm{L}}\, P_{\mathrm{L}} P_{\mathrm{L}}^{\mathsf{T}},
\qquad \alpha_{\mathrm{S}}, \alpha_{\mathrm{L}} \ge 0,
\label{eq:C-safe}
\end{equation}
which is positive semidefinite by construction for any real
$P_{\mathrm{S}}, P_{\mathrm{L}}$ and is the appropriate choice
whenever the modes are treated as \emph{basis operators} (outputs
of an orthogonalization procedure) rather than as
already-validated covariance matrices. Note that \eqref{eq:C-safe}
constrains the cross-scale covariance to zero by construction; where
a cross-scale term is itself of interest, form (i) is the appropriate
tool. Assembly (ii) is implemented
(non-negative least-squares amplitudes on the per-scale squared
operator modes plus nugget) and exercised on the real posterior and in
the Monte Carlo study of Sect.~\ref{sec:diagresults}.

\paragraph{(iii) Direct amplitude form.} The direct combination
\begin{equation}
C(\alpha_{\mathrm{S}}, \alpha_{\mathrm{L}})
 = \alpha_{\mathrm{S}}\, P_{\mathrm{S}}
 + \alpha_{\mathrm{L}}\, P_{\mathrm{L}}
\label{eq:C-direct}
\end{equation}
is positive semidefinite only if the modes themselves are; the
Gram--Schmidt step \eqref{eq:gsop2} subtracts a multiple of $P_1$
from $\KL$ and can introduce negative eigenvalues in $P_2$ even when
both basis functions are individually admissible covariance matrices, so
\eqref{eq:C-direct} may be used only after an explicit eigenvalue
check. In forms (ii) and (iii) a nugget term $\alpha_0 I$ (local
noise; the ``nugget effect'' of geostatistics, \citealt{cressie1993})
is typically added, and positive semidefiniteness of the final
operator must be verified numerically, not assumed.
Assembly (iii), including the nugget term, the explicit
eigenvalue check, and a nugget repair when the check fails, is
likewise implemented and exercised in
Sect.~\ref{sec:diagresults}.

Orthogonality in \eqref{eq:orthonormal} is relative to $W$. Changing
$W$ (e.g., between \wFBET{} iterations) changes the modes. If
$W$ is updated iteratively, the orthogonalization must be repeated per
iteration, and convergence of the joint loop (weights $\to$ modes
$\to$ fit $\to$ weights) should be monitored rather than presumed.

\section{Results}
\label{sec:results}

This section reports the weighted Gram--Schmidt bookkeeping
itself: the constructed modes, the per-scale and cross-scale outputs on
the two datasets, and the dependence on the two implemented
hyperparameter-cost conventions. The predeclared diagnostics and the
comparisons against alternative constructions are collected in the
Discussion (Sect.~\ref{sec:discussion}).
\subsection{Mode construction and orthogonalization}

Basis-function families of equal size per scale are used:
$N_{\mathrm{S}} = N_{\mathrm{L}} = n_m$, with $n_m = 5$ (ten
basis functions on the eleven-node grid, close to saturation) for
AirPassengers and $n_m = 7$ for the luminosity case (fourteen basis
functions on the $n_g = 14$ grid). The single-index
realization \eqref{eq:single-short}--\eqref{eq:single-long} is
evaluated on the grid: seasonal bumps
$\kS(\dS(\xi, m_a);\, \LS^{(a)})$, with wrapped distance for
AirPassengers and unwrapped distance for the luminosity case, and
long profiles $\kL(\xi;\, \muL^{(b)}, \sigL^{(b)})$ anchored at the
series origin. Each basis function vector is normalized to unit Euclidean
norm, the family is stacked short-first
(the declared short-first ordering), and the stack is orthogonalized by an
economy QR factorization under the declared baseline metric $W = I$
(Section~\ref{sec:versionA}). The weighted Gram--Schmidt recursion is
the drop-in generalization, and variation of $W$ is among the
robustness checks below.

\subsection{Bookkeeping outputs}

\begin{figure*}[!tp]
\centering
\includegraphics[width=0.95\textwidth]{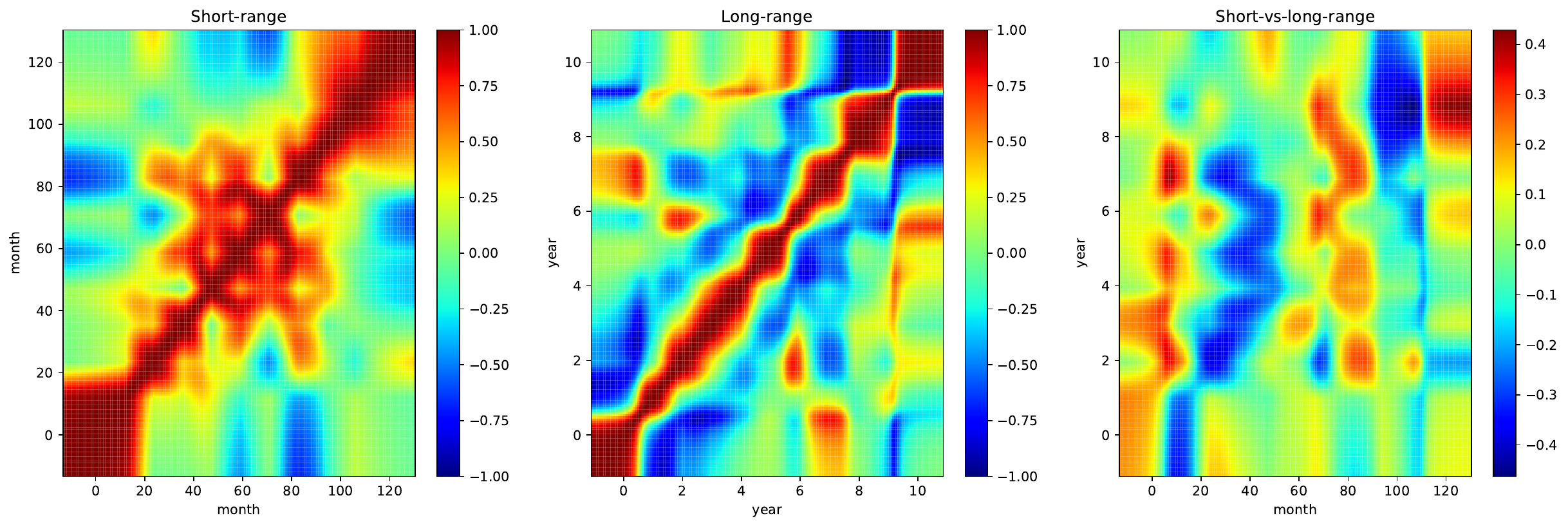}
\caption{AirPassengers per-scale bookkeeping blocks from the
implemented run, shown as correlation matrices on the prediction grid
with the scale labels inherited from the basis function families.
\textit{Left:} short-range block
$\widehat{C}_{\mathcal{S}\mathcal{S}}$ over the month coordinate.
\textit{Middle:} long-range block
$\widehat{C}_{\mathcal{L}\mathcal{L}}$ over the chronological
coordinate (years). \textit{Right:} cross-scale block
$\widehat{C}_{\mathcal{S}\mathcal{L}}$ (months versus years); a
nonzero cross block is a reported finding, not an error
(Sect.~\ref{sec:assembly}). Coordinates are shifted to zero minimum
(Sect.~\ref{sec:prototype}); color bars give the correlation value.
Run with the projection-fidelity cost \eqref{eq:projcost};
cross-block correlations lie within $\pm0.35$.}
\label{fig:airpassengers-long-short}
\end{figure*}

\begin{figure*}[!tp]
\centering
\includegraphics[width=0.95\textwidth]{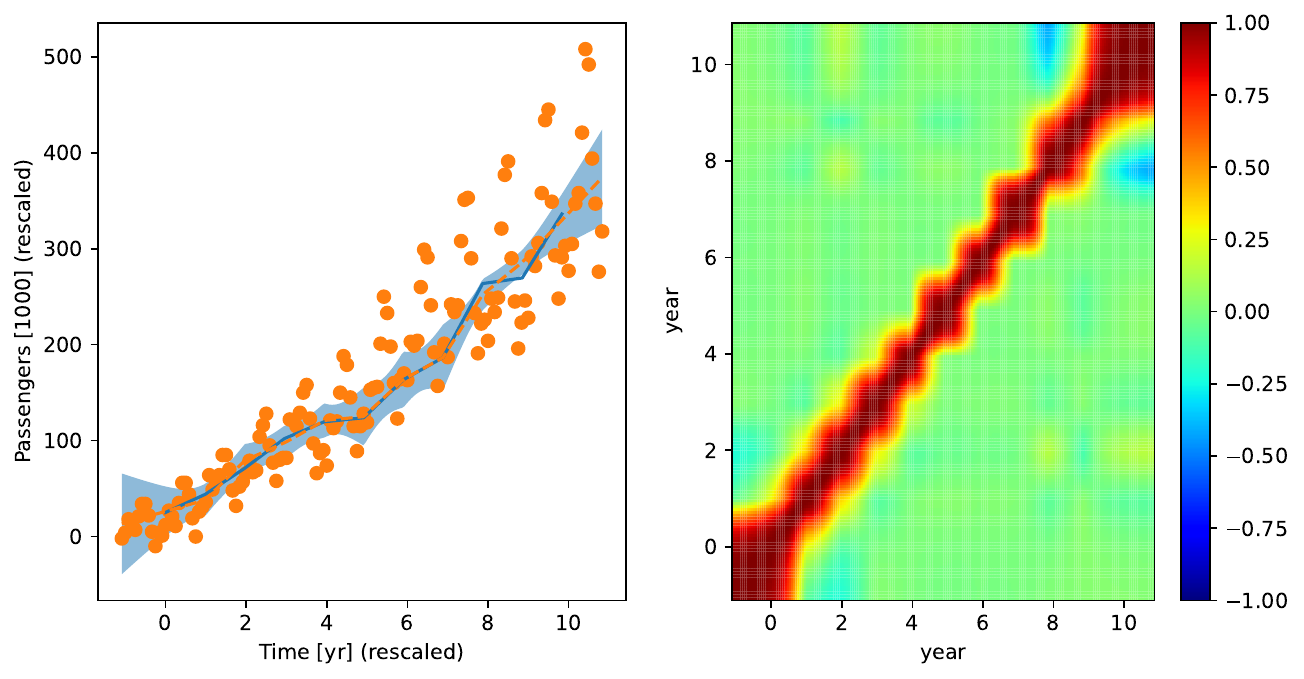}
\caption{AirPassengers posterior reprojection diagnostic. The FBET
posterior mean and covariance are projected into the selected
finite-rank bookkeeping basis and reconstructed on the prediction
grid. \textit{Left:} reconstruction $\widehat{x}$ (dashed) with its
one-standard-deviation band, overlaid on the shifted data (points),
including the held-out 1960 segment. \textit{Right:} correlation
structure of the reconstructed covariance $\widehat{C}$ over the
chronological coordinate (years). Together the panels give a direct
visual check of the information retained by the selected modes.
Run with the projection-fidelity cost \eqref{eq:projcost}.
Data and reconstruction share a common (training-window)
origin; points preceding the training window appear at negative
abscissae.}
\label{fig:airpassengers-reproj}
\end{figure*}

\begin{figure*}[!tp]
\centering
\includegraphics[width=0.95\textwidth]{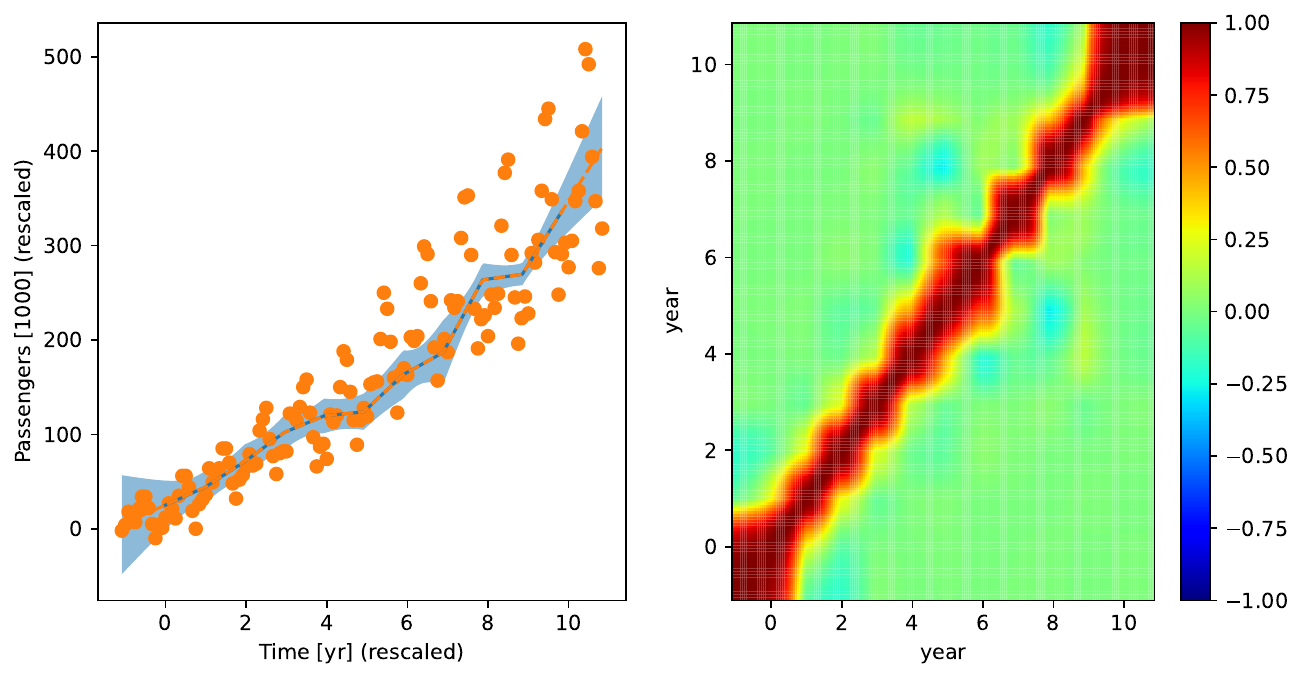}
\caption{Same as Fig.~\ref{fig:airpassengers-reproj}
(AirPassengers posterior reprojection), for the
pseudo-inverse--weighted cost variant of Sect.~\ref{sec:hyperopt}.
The reconstructed correlation structure differs in the off-diagonal
(anticorrelated) regions, again reflecting the cost-convention
dependence of the selected hyperparameters.}
\label{fig:airpassengers-reproj-alt}
\end{figure*}

\begin{figure*}[!tp]
\centering
\includegraphics[width=0.95\textwidth]{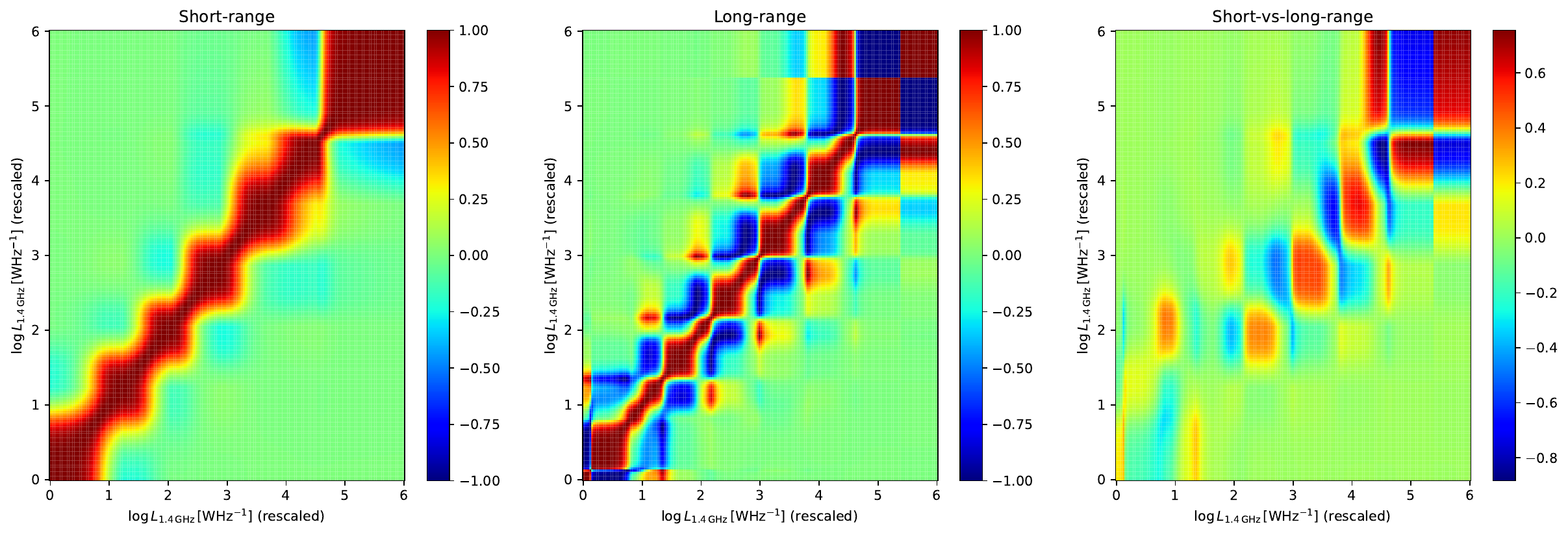}
\caption{Luminosity-function per-scale bookkeeping blocks, in the
same arrangement as Fig.~\ref{fig:airpassengers-long-short}
(\textit{left:} $\widehat{C}_{\mathcal{S}\mathcal{S}}$;
\textit{middle:} $\widehat{C}_{\mathcal{L}\mathcal{L}}$;
\textit{right:} $\widehat{C}_{\mathcal{S}\mathcal{L}}$). In this
non-cyclic test case the short-scale basis function uses
unwrapped distance, while the channel-weighted FBET stage exercises
the interaction between marginalized channel weighting and the orthogonalized
correlation-scale decomposition.}
\label{fig:lf-long-short}
\end{figure*}

The selected modes and projections yield: the reconstruction
$(\widehat{x}, \widehat{C})$ on a finer prediction grid, obtained by
interpolating the per-mode contributions; the per-scale covariance
blocks $\widehat{C}_{\mathcal{S}\mathcal{S}}$ and
$\widehat{C}_{\mathcal{L}\mathcal{L}}$ and the cross block
$\widehat{C}_{\mathcal{S}\mathcal{L}}$ of \eqref{eq:blocks},
displayed as correlation matrices over the seasonal and chronological
coordinates respectively
(Figs.~\ref{fig:airpassengers-long-short}
and~\ref{fig:lf-long-short} for the two test cases); and, for
AirPassengers, an overlay of the
reconstruction and its one-standard-deviation band on the held-out
data on both sides of the training window
(Fig.~\ref{fig:airpassengers-reproj}): the 1960 segment above it and
the January 1949--January 1950 segment below it, the latter appearing
at negative abscissae. This two-sided overlay is the purpose of the
deliberately narrowed training window of
Sect.~\ref{sec:prototype}, it shows directly whether the
extrapolated reconstruction and its uncertainty band track the real
observations beyond the fitted interval in either direction.
The data overlay and the reconstruction share a common
origin (both are shifted by the training-window minima), so the
holdout comparison in Figs.~\ref{fig:airpassengers-reproj}
and~\ref{fig:airpassengers-reproj-alt} is directly readable. The
prediction-grid reconstruction extrapolates the
fitted modes beyond the training window; this is an inspection
device, and no claim is made that the finite-rank basis is an
adequate forecasting model.

\begin{figure*}[!tp]
\centering
\includegraphics[width=0.95\textwidth]{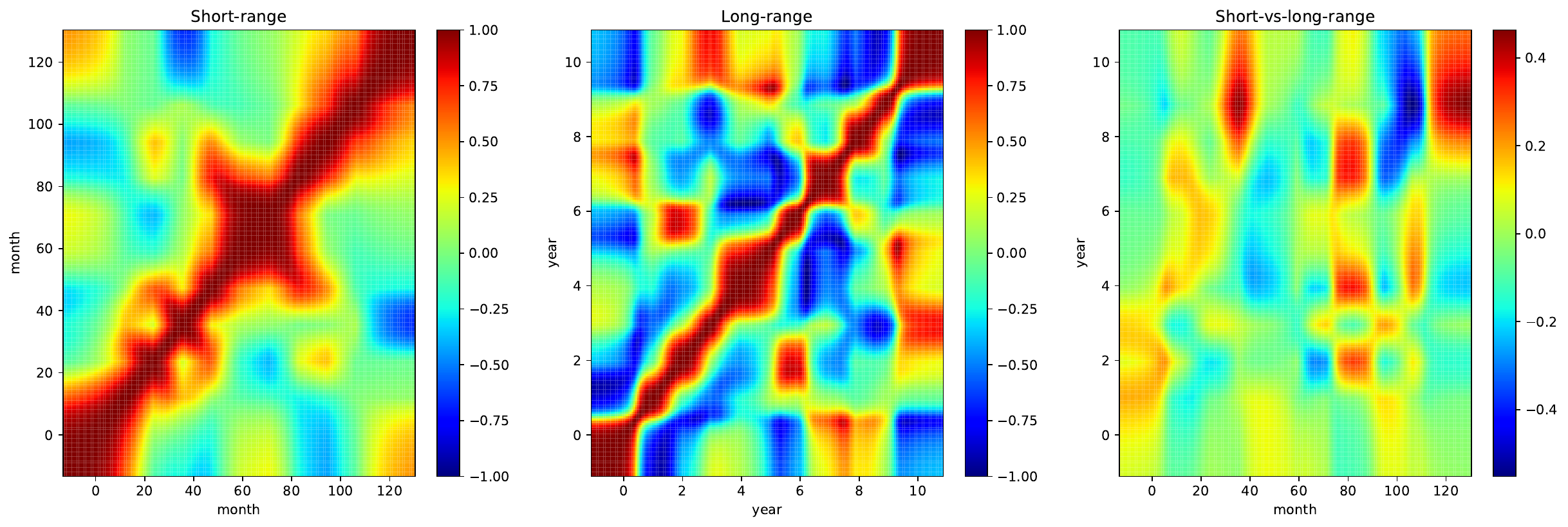}
\caption{Same as Fig.~\ref{fig:airpassengers-long-short}
(AirPassengers per-scale blocks), for the pseudo-inverse--weighted
cost variant of Sect.~\ref{sec:hyperopt}. The selected hyperparameters, and hence the
per-scale blocks, differ visibly from
Fig.~\ref{fig:airpassengers-long-short} (cross-block correlations
within $\pm0.45$ and a different sign pattern), illustrating the
declared-convention dependence quantified in
Sect.~\ref{sec:diagresults}.}
\label{fig:airpassengers-long-short-alt}
\end{figure*}

\subsection{Per-run diagnostics and cost-convention dependence}

\begin{figure*}[!tp]
\centering
\includegraphics[width=0.95\textwidth]{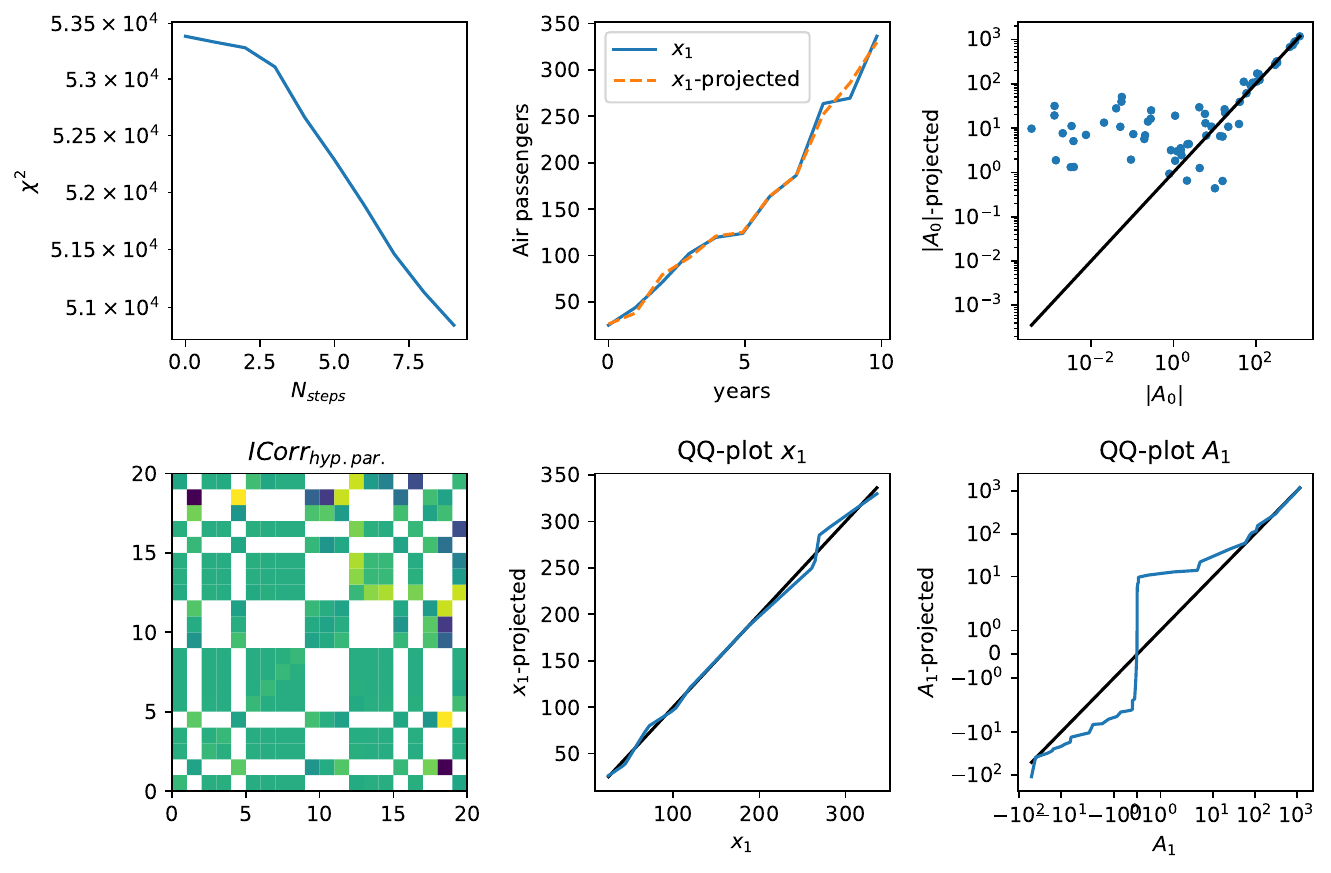}
\caption{AirPassengers diagnostic summary for the implemented run,
collecting the predeclared checks of Sect.~\ref{sec:diagnostics}.
\textit{Top row:} trace of the cost \eqref{eq:projcost} over search
iterations; posterior mean versus its projection; element-wise
comparison of posterior and reconstructed covariance magnitudes on a
log--log identity plot. \textit{Bottom row:} hyperparameter
correlation matrix from the final Hessian; quantile--quantile
comparisons of the posterior mean and covariance against their
reconstructions. (The magnitude-panel labels $|A_0|$ refer to the
posterior covariance denoted $A_1$ in the text.)
Run with the projection-fidelity cost \eqref{eq:projcost}.}
\label{fig:airpassengers-diagnostics}
\end{figure*}

\begin{figure*}[!tp]
\centering
\includegraphics[width=0.95\textwidth]{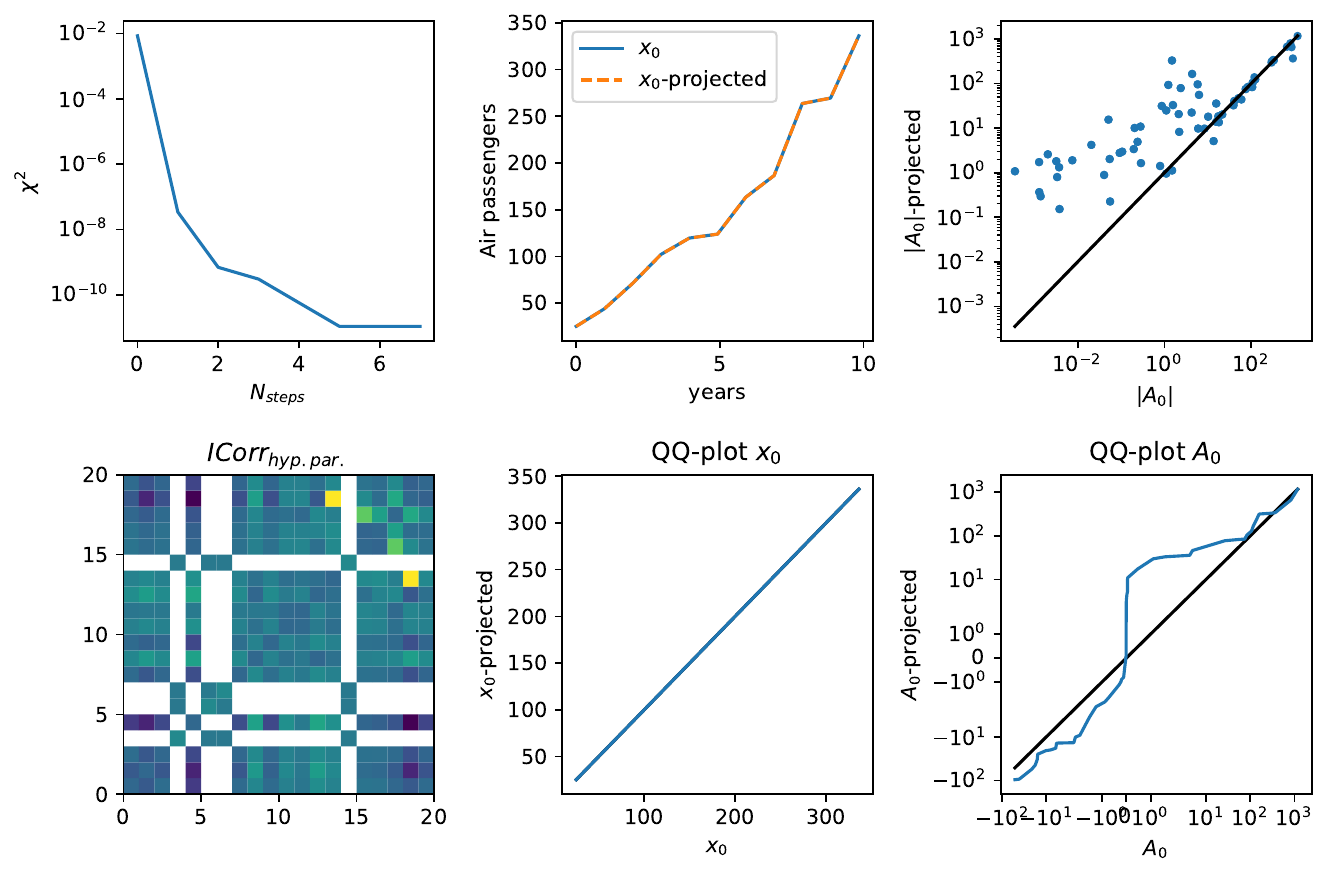}
\caption{Same as Fig.~\ref{fig:airpassengers-diagnostics}
(AirPassengers diagnostic summary), for the pseudo-inverse--weighted
cost variant of Sect.~\ref{sec:hyperopt}; note the different (Mahalanobis-normalized)
scale of the cost trace.}
\label{fig:airpassengers-diagnostics-alt}
\end{figure*}

Figures~\ref{fig:airpassengers-diagnostics}
and~\ref{fig:airpassengers-diagnostics-alt} collect the per-run
diagnostics for the two implemented cost conventions. Under the
projection-fidelity cost \eqref{eq:projcost} the cost trace declines smoothly from $5.35\times10^{4}$
to $\approx5.05\times10^{4}$ over the accepted steps; the
pseudo-inverse--weighted variant behaves comparably on its own
(differently normalized) scale. In both runs the
quantile--quantile comparison of the posterior mean against its
projection is close to the identity, while the covariance
comparisons show the expected signature of a finite-rank
compression: large-magnitude elements of $A_1$ follow the identity
line, small-magnitude elements are over-represented by the
reconstruction, and the covariance quantile--quantile curve departs
from the identity in the mid-range. Masked (white) entries of the
hyperparameter correlation matrices correspond to entries left
undefined by the pseudo-inverse of the final Hessian on the search
grid. The two cost conventions select visibly different per-scale
blocks (Figs.~\ref{fig:airpassengers-long-short}
and~\ref{fig:airpassengers-long-short-alt}: cross-block correlations
within $\pm0.35$ versus $\pm0.45$, with different sign patterns) and
different reconstructed correlation structures
(Figs.~\ref{fig:airpassengers-reproj}
and~\ref{fig:airpassengers-reproj-alt}); the choice of
projection-fidelity cost is therefore a consequential declared
convention rather than a numerical detail. In the extrapolated region
of the prediction grid (beyond $\approx10$~yr) the finite-rank
reconstruction saturates towards full correlation, underlining that
the prediction-grid extension is an inspection device, not a
forecasting model.

\begin{figure*}[!tp]
\centering
\includegraphics[width=0.95\textwidth]{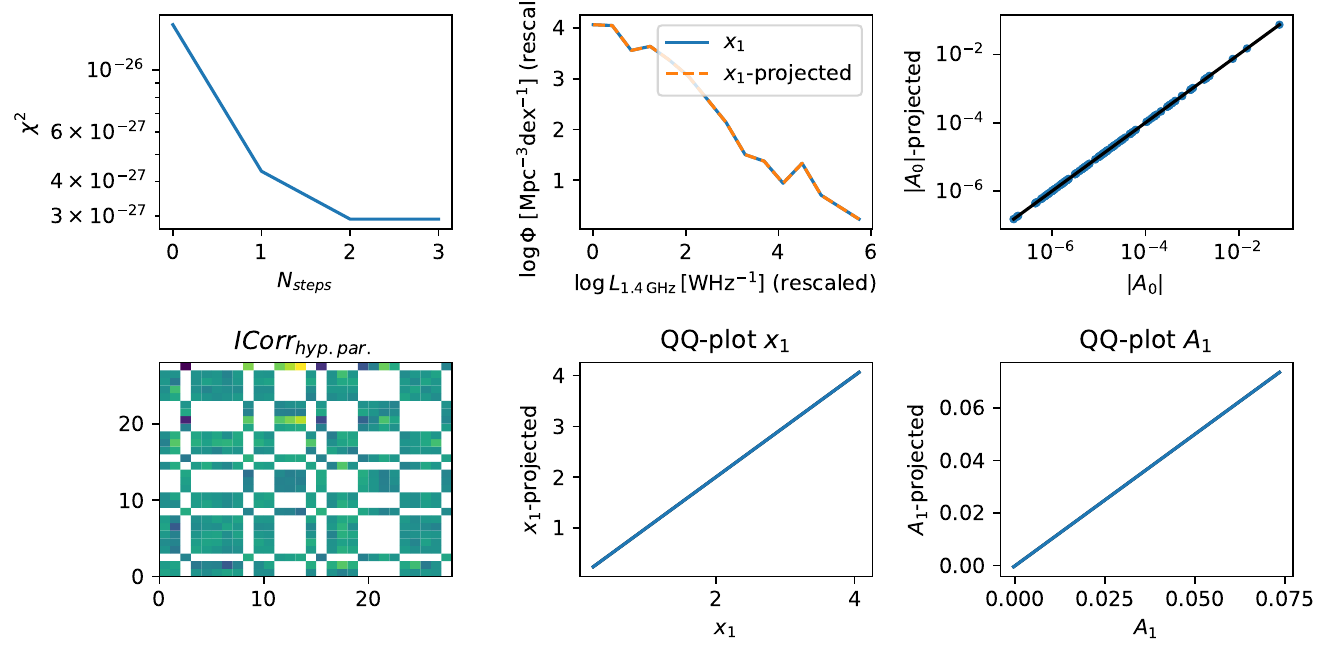}
\caption{Luminosity-function diagnostic summary for the weighted
implemented run, in the same arrangement as
Fig.~\ref{fig:airpassengers-diagnostics}. The diagnostics mirror the
AirPassengers checks while
also reflecting the channel-weighted covariance update used for the
redshift-bin grouped data. (The magnitude-panel labels $|A_0|$ refer to the
posterior covariance denoted $A_1$ in the text.)}
\label{fig:lf-diagnostics}
\end{figure*}
\subsection{The luminosity-function run}
\label{sec:lfresults}

\begin{figure*}[!tp]
\centering
\includegraphics[width=0.95\textwidth]{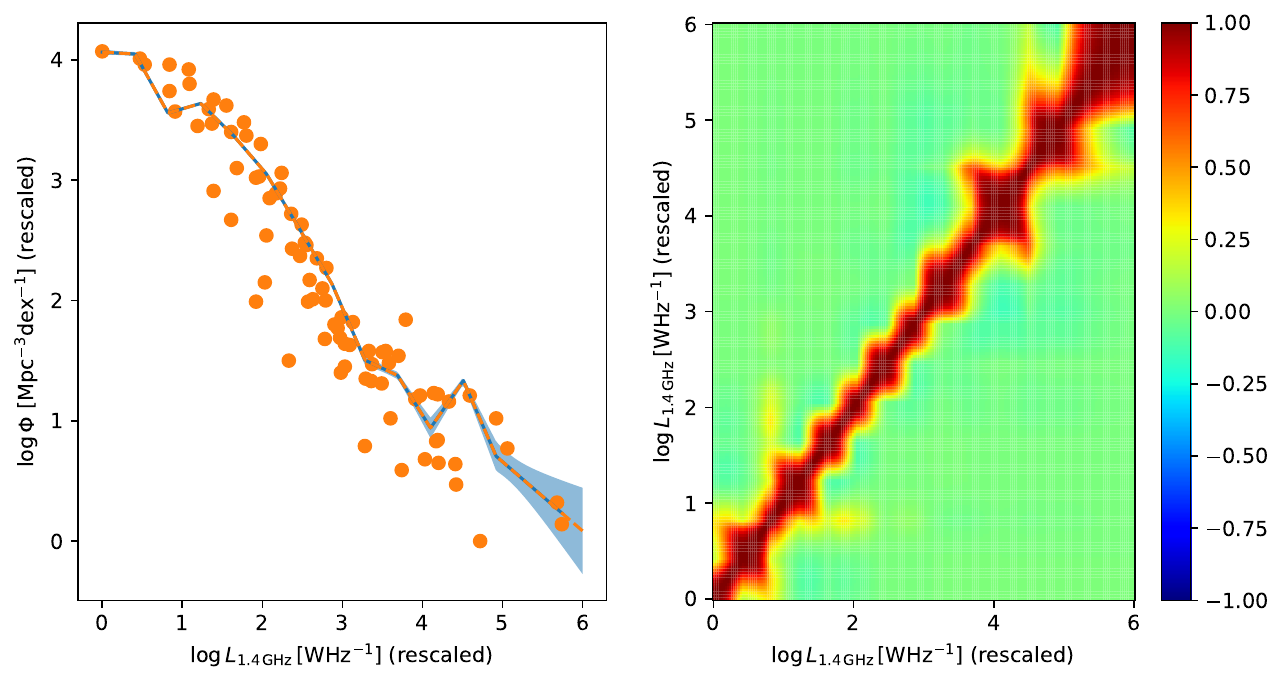}
\caption{Luminosity-function posterior reprojection diagnostic, in
the same arrangement as Fig.~\ref{fig:airpassengers-reproj}, for the
pseudo-inverse--weighted cost variant of Sect.~\ref{sec:hyperopt}.
\textit{Left:} the preprocessed measurements (points; the ordinate
is $\log\Phi$, shifted to zero minimum)
with the reconstruction $\widehat{x}$ (dashed) and its
one-standard-deviation band on the prediction grid. \textit{Right:}
correlation structure of the reconstructed covariance over the
shifted luminosity coordinate.}
\label{fig:lf-reproj}
\end{figure*}

The same diagnostic suite, adapted to the non-cyclic abscissa
(unwrapped short-range distance, residual structure reported against
the coordinate and against the redshift bin instead of the
seasonal phase) and to the weighted FBET stage with Poisson channel
weights, was executed for the luminosity-function case. The pipeline
configuration uses $n_m = 7$, i.e.\ fourteen basis functions on the
$n_g = 14$ evaluation grid; the dictionary is square (complete), all
fourteen modes survive the rank-aware orthogonalization, and the
first-seven/last-seven scale partition inherited by the block
bookkeeping is faithful. As for AirPassengers, the projection
diagnostics are executed in two metric variants, the declared
baseline $W=I$ and the posterior precision
$W = A_1^{+}$. The reprojection diagnostic for this
run is shown in Fig.~\ref{fig:lf-reproj}, complementing the
per-scale blocks of Fig.~\ref{fig:lf-long-short} and the per-run
summary of Fig.~\ref{fig:lf-diagnostics}.

\section{Discussion}
\label{sec:discussion}

We now turn to the predeclared diagnostic suite and to the
comparisons against alternative constructions (the naive additive
dictionary, the operator-valued realization, and the amplitude
assemblies) which contextualize the results of
Sect.~\ref{sec:results}.

\subsection{Why orthogonalization is needed}
\label{sec:whygs}

Consider the naive additive dictionary model
\begin{equation}
\begin{aligned}
C_{\mathrm{naive}}
 &= \alpha_{\mathrm{S}}\, \KS + \alpha_{\mathrm{L}}\, \KL,\\
[\KS]_{ij} &= \kS(i,j;\, \LS),\\
[\KL]_{ij} &= \kL(i,j;\, \muL, \sigL).
\end{aligned}
\label{eq:naive}
\end{equation}
We state explicitly: \eqref{eq:naive} is only a
\emph{pre-orthogonalized dictionary representation}. It is admissible
as a modeling ansatz if and only if it is understood (and declared) as such, and it can double-count correlation structure unless
$\KS$ and $\KL$ have been made orthogonal under the chosen metric.
The same warning applies to the covariance-level shorthand
$\Sigma = \Sigma_{\mathrm{short}} + \Sigma_{\mathrm{long}}$, which we
avoid except where explicitly qualified in this way. The practical
failure modes of \eqref{eq:naive} are the ones sketched in
Section~\ref{sec:infogeo}:
\begin{itemize}[itemsep=2pt]
\item \emph{Double counting.} Structure lying in the overlap of the
two basis functions is charged to both amplitudes, inflating the total
implied correlation.
\item \emph{Loss of interpretability.} The fitted
$(\alpha_{\mathrm{S}}, \alpha_{\mathrm{L}})$ are not attributable to
``seasonal'' and ``long-range'' effects separately, because each
basis function partially contains the other.
\item \emph{Poor identifiability.} The larger the overlap, the closer
to singular the hyperparameter estimation problem for
$(\alpha_{\mathrm{S}}, \alpha_{\mathrm{L}})$ becomes.
\item \emph{Metric instability.} Downstream, $W = C^{-1}$ (where
defined) and the induced FIM \eqref{eq:fim-gls} inherit the
ill-conditioning.
\item \emph{Misleading uncertainty attribution.} Propagated
uncertainty budgets that report per-scale contributions become
parametrization artifacts.
\end{itemize}
Weighted Gram--Schmidt orthogonalization of the basis functions, under an
explicitly chosen positive-definite metric, addresses all five points
at the cost of making the resulting modes metric-dependent, a cost
we return to in Section~\ref{sec:discussion}.

\subsection{Predeclared diagnostics}
\label{sec:diagnostics}

Implemented diagnostics, reported per run
(Figs.~\ref{fig:airpassengers-diagnostics}
and~\ref{fig:lf-diagnostics} for the two test cases): the trace of
the cost
\eqref{eq:projcost} over search iterations; the hyperparameter
correlation matrix from the final Hessian; an element-wise comparison
of $\lvert A_1 \rvert$ against $\lvert \widehat{C} \rvert$ on a
log--log identity plot; quantile--quantile comparisons of $x_1$
versus $\widehat{x}$ and of $A_1$ versus $\widehat{C}$; and the
holdout overlay described above.

The following checks were fixed in advance and executed for the
AirPassengers case (results in
Sect.~\ref{sec:diagresults}): the Gram matrix
\begin{equation}
G_{ab} = \ip{\psi_a}{\psi_b},
\label{eq:gram}
\end{equation}
with $\lVert G - I \rVert$ reported, trivially satisfied under the
QR construction with $W = I$, but a required check for every weighted
variant; the \emph{pre-orthogonalization} overlap matrix (the
normalized Gram matrix of the raw basis functions), which measures
directly how much double counting the naive dictionary
\eqref{eq:naive} would have committed; and stability of hyperparameters
and per-scale blocks under (i) hyperparameter perturbation, (ii) swap
of the orthogonalization ordering, and (iii) change of $W$ among
declared variants. The same checks, adapted to the non-cyclic
abscissa and the weighted FBET stage, are reported for the
luminosity-function case in Sect.~\ref{sec:lfresults}.

The expected qualitative outcomes (stated as hypotheses to be
checked, not as results) are that the raw short- and long-range
basis function families overlap materially (pairs at integer-year
separations are close in $\dS$ while well separated in $\dL$, and the
single-index seasonal bumps are not mutually orthogonal); that the
cross-scale block $\widehat{C}_{\mathcal{S}\mathcal{L}}$ is nonzero
and interpretable; and that the naive dictionary shows degraded
identifiability relative to the orthogonalized bookkeeping.
All three hypotheses are confirmed quantitatively in
Sect.~\ref{sec:diagresults}.

\subsection{Diagnostics for AirPassengers}
\label{sec:diagresults}

We now report the outcome of the predeclared checks for the
AirPassengers case. All quantities in this subsection are produced by
the implementation at the selected hyperparameters; unless stated
otherwise, the run uses the projection-fidelity cost
\eqref{eq:projcost} and the declared baseline metric $W=I$ on the
evaluation grid.

\begin{figure*}[!tp]
\centering
\includegraphics[width=0.95\textwidth]{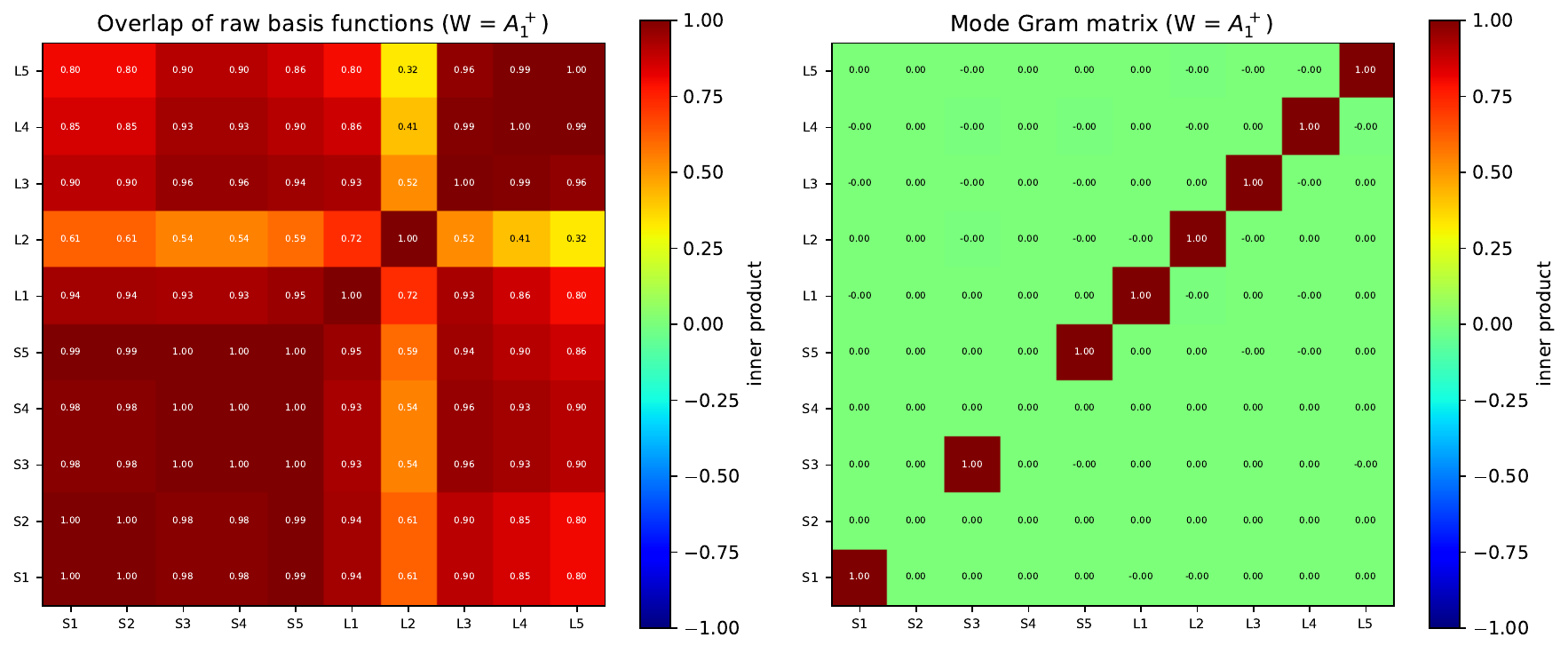}
\caption{Double-counting diagnostic for AirPassengers under
the posterior-precision metric $W = A_1^{+}$.
\textit{Left:} pre-orthogonalization overlap matrix (normalized Gram
matrix) of the ten raw basis functions of the naive dictionary
\eqref{eq:naive} (five seasonal bumps (S1--S5) and five long-range
profiles (L1--L5)) on the evaluation grid; all pairwise overlaps
lie between $0.32$ and $1.00$. \textit{Right:} Gram matrix
$G_{ab}=\langle\psi_a,\psi_b\rangle_W$ of the orthonormalized modes
(Eqs.~\ref{eq:orthonormal} and~\ref{eq:gram}), equal to the identity
to machine precision; under this metric the projected mode-space
covariance is itself the identity (whitening property,
Sect.~\ref{sec:diagresults}). The identity-metric variant is shown in
Fig.~\ref{fig:gram-overlap} (Appendix~\ref{app:identity}).}
\label{fig:gram-overlap-winv}
\end{figure*}

\paragraph{Pre-orthogonalization overlap and mode Gram
matrix.}
Figure~\ref{fig:gram-overlap-winv} shows the normalized Gram
(overlap) matrix of the ten raw basis functions next to the Gram matrix
\eqref{eq:gram} of the orthonormalized modes, under the
posterior-precision metric $W = A_1^{+}$. The
overlap is severe:
the overlaps lie between $\approx0.32$ and $1.00$ with the
large majority above $0.8$; the five seasonal bumps are pairwise
indistinguishable at the quoted precision (overlap $1.00$), and the
short--long cross overlaps reach $\approx0.95$. The naive dictionary
\eqref{eq:naive} would therefore double-count essentially all of its
structure, and its Gram matrix is nearly singular, a direct,
quantitative confirmation of the double-counting failure mode of
Sect.~\ref{sec:whygs}. After orthogonalization the mode Gram matrix equals
the identity to machine precision, verifying \eqref{eq:orthonormal}
under this metric. Under it, moreover, the projected mode-space
covariance is the identity exactly,
$A_{\mathrm{proj}} = \Psi W A_1 W \Psi^{\mathsf T}
 = \Psi W \Psi^{\mathsf T} = I$
on the surviving modes, so the cross-scale block vanishes by
construction and the per-scale attribution becomes purely geometric
(determined by the spans of the mode families alone), a whitening
property that makes the posterior-precision metric the statistically
natural, but attribution-trivializing, choice.
The identity-metric baseline
(Fig.~\ref{fig:gram-overlap}, Appendix~\ref{app:identity}) shows a
nearly identical overlap pattern (range $0.34$--$1.00$), so the
double-counting risk is a property of the family design rather than
of the declared metric. The near-coincidence of the
seasonal bumps also signals that the selected basis function family itself
is close to degenerate; a smaller seasonal family, or constrained
reference phases, would be preferable and is left for the next
iteration of the protocol.

\begin{figure*}[!tp]
\centering
\includegraphics[width=0.9\textwidth]{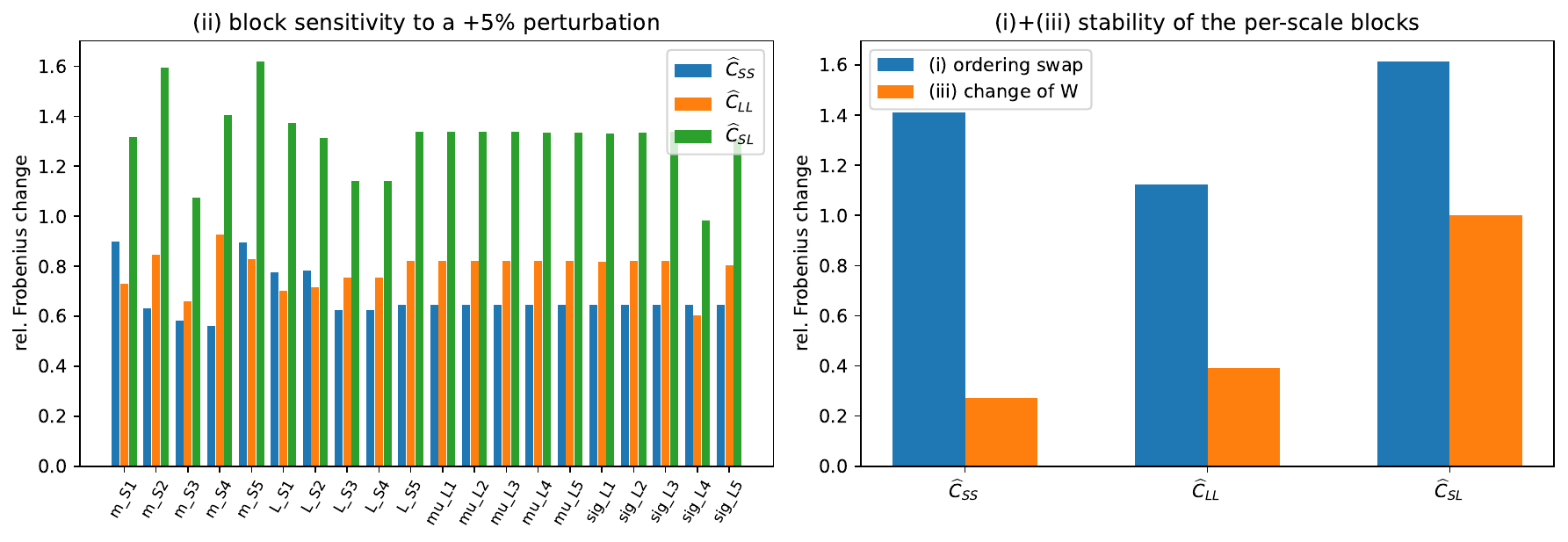}
\caption{Stability of the per-scale blocks \eqref{eq:blocks}
for AirPassengers. \textit{Left:} relative Frobenius change of
$\widehat{C}_{\mathcal{S}\mathcal{S}}$,
$\widehat{C}_{\mathcal{L}\mathcal{L}}$, and
$\widehat{C}_{\mathcal{S}\mathcal{L}}$ under a $+5\%$ perturbation of
each basis function hyperparameter. \textit{Right:} changes under a swap
of the orthogonalization ordering and under a change of the declared
metric $W$ (from $W=I$ to a diagonal-precision metric). The total
reconstruction $\widehat{C}$ is invariant under the ordering swap to
machine precision; the individual blocks are not.}
\label{fig:robustness}
\end{figure*}

\paragraph{Robustness of the per-scale blocks.}
Figure~\ref{fig:robustness} quantifies the stability of the
per-scale blocks \eqref{eq:blocks} under the three predeclared
variations. Swapping the orthogonalization ordering leaves the total
reconstruction $\widehat{C}$ invariant to machine precision (the
spanned subspace does not depend on the order) but changes the
individual blocks by relative Frobenius differences of order unity,
and replacing the baseline metric $W=I$ by a diagonal-precision
metric changes them by factors ranging from $\approx6$
($\widehat{C}_{\mathcal{S}\mathcal{S}}$) and $\approx7.5$
($\widehat{C}_{\mathcal{L}\mathcal{L}}$) to $\approx21$
($\widehat{C}_{\mathcal{S}\mathcal{L}}$). A $+5\%$ perturbation of
individual hyperparameters shows the blocks to be most sensitive to the seasonal hyperparameters
(relative changes up to $\approx0.9$) and to a subset of the
long-range widths (up to $\approx0.5$), with the remaining
long-range hyperparameters nearly inert. Taken together with the
near-degenerate overlaps of Fig.~\ref{fig:gram-overlap-winv}, the
conclusion is that for this fitted basis function family the per-scale
attribution is strongly convention-dependent, exactly as anticipated
in Sects.~\ref{sec:whygs} and~\ref{sec:assembly}: any reported
per-scale block must carry the declared ordering, metric, and
hyperparameters alongside it.

\begin{figure}[!tp]
\centering
\includegraphics[width=\columnwidth]{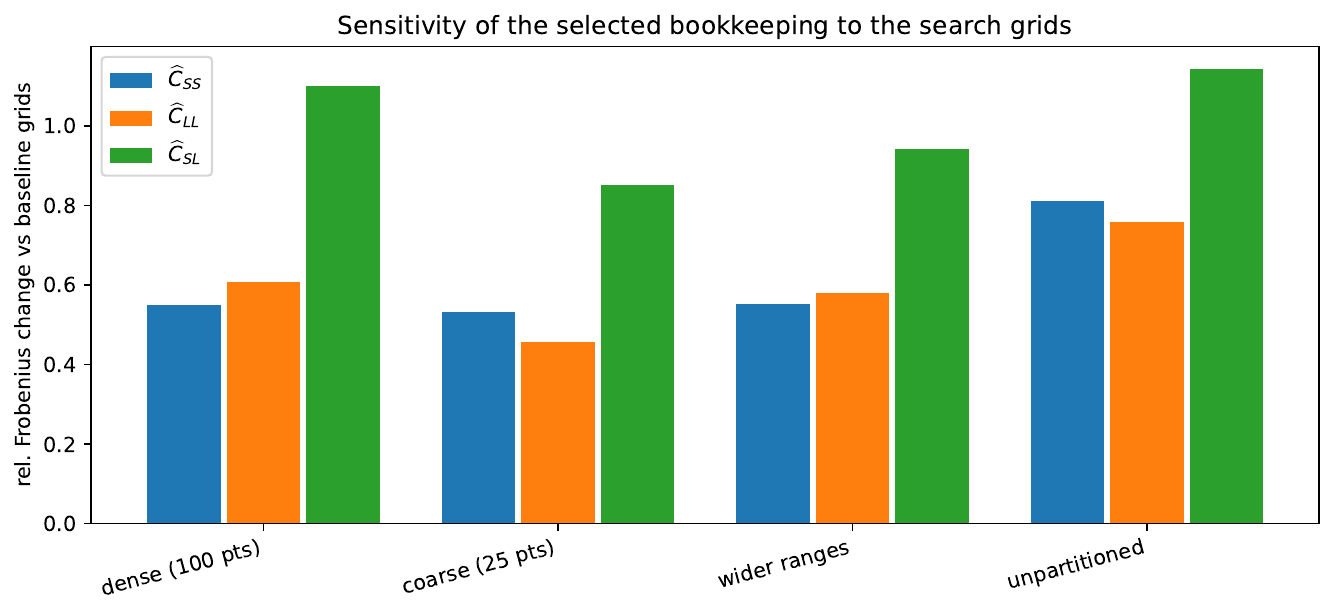}
\caption{Sensitivity of the selected AirPassengers
bookkeeping to the declared search grids
(Sect.~\ref{sec:hyperopt}): relative Frobenius change of the
per-scale blocks $\widehat{C}_{\mathcal{S}\mathcal{S}}$,
$\widehat{C}_{\mathcal{L}\mathcal{L}}$, and
$\widehat{C}_{\mathcal{S}\mathcal{L}}$ against the baseline grids,
for the four predeclared grid variants: doubled density (100
points), halved density (25 points), widened $\LS$ and $\sigL$
ranges, and unpartitioned anchor grids.}
\label{fig:grid-sensitivity}
\end{figure}

\paragraph{Grid sensitivity.}
Figure~\ref{fig:grid-sensitivity} reports the predeclared
grid-sensitivity protocol of Sect.~\ref{sec:hyperopt}: the
hyperparameter search is rerun with doubled (100-point) and halved
(25-point) grid density, with widened $\LS$ and $\sigL$ ranges, and
with unpartitioned anchor grids, and the per-scale blocks of the
reselected bookkeeping are compared with the baseline. The changes
are large: the relative Frobenius change of the short- and
long-range blocks lies between $\approx0.45$ and $\approx0.8$ across
the four variants, and that of the cross block between $\approx0.85$
and $\approx1.15$, with the largest changes under the unpartitioned
anchor grids, the variant that restores the permutation freedom
the partition was introduced to remove. The selected hyperparameters
therefore depend materially on the declared search grids, and the
grid specification of Sect.~\ref{sec:hyperopt} must be treated as
part of the bookkeeping convention, on the same footing as the
metric $W$, the orthogonalization ordering, and the cost.

\begin{figure*}[!tp]
\centering
\includegraphics[width=0.95\textwidth]{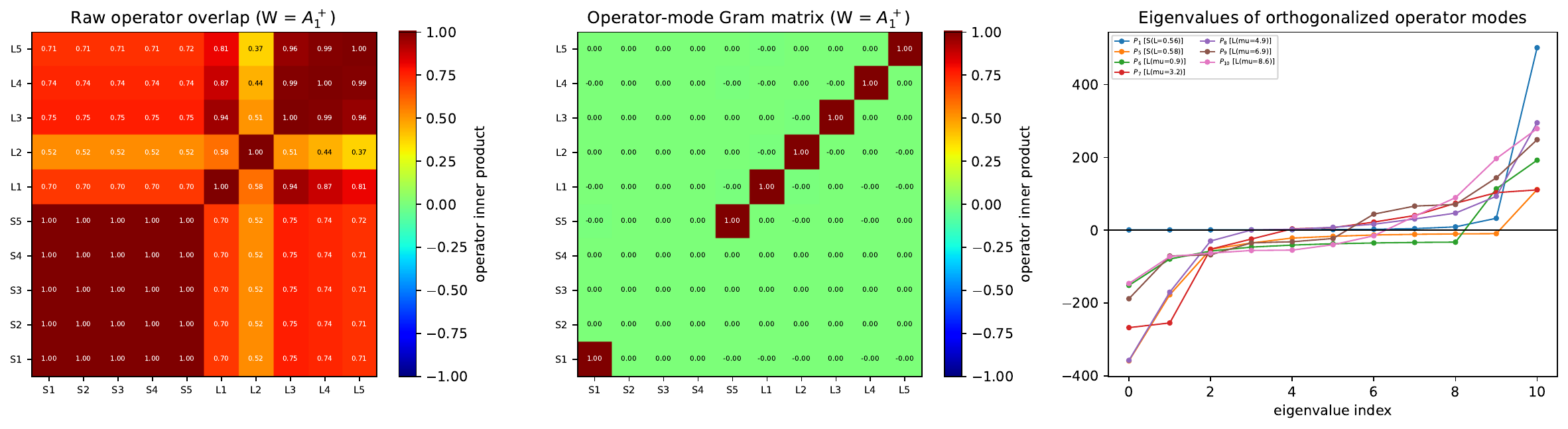}
\caption{Operator-valued (Version B,
Sect.~\ref{sec:versionB}) check for AirPassengers under the
posterior-precision metric $W = A_1^{+}$.
\textit{Left:} normalized operator Gram (overlap) matrix of the ten
two-index kernels on the evaluation grid, under the operator inner
product \eqref{eq:ipop}; the five short-range kernels are
operator-identical at the quoted precision. \textit{Middle:}
operator-mode Gram matrix after two Gram--Schmidt passes, equal to
the identity on the surviving modes. \textit{Right:} eigenvalue
spectra of the surviving modes: only the leading mode is positive
semidefinite; the magnitudes reach a few hundred because the modes
are normalized in the weighted operator norm. The identity-metric
variant is shown in Fig.~\ref{fig:operator-gs}
(Appendix~\ref{app:identity}).}
\label{fig:operator-gs-winv}
\end{figure*}

\paragraph{Operator-valued (Version B) check.}
Figure~\ref{fig:operator-gs-winv} reports the operator-space
analogue of Fig.~\ref{fig:gram-overlap-winv}, built from the two-index
kernels on the evaluation grid and the operator inner product
\eqref{eq:ipop}, again under the posterior-precision
metric. The raw operator overlaps repeat the vector-space
picture in sharpened form: the five short-range kernels, which
in this realization depend on the correlation length alone, are
operator-identical at the quoted precision (mutual overlaps $1.00$),
two of the five long-range members nearly coincide, and seven of the
ten operator modes survive the rank-aware orthogonalization
(degenerate directions are kept as zero slots). On the surviving
modes the operator Gram matrix equals the identity to machine
precision. The right panel shows the per-mode eigenvalue spectra: the
first mode, a rescaled admissible kernel, is positive semidefinite,
while every subsequent surviving mode is indefinite; the
magnitudes reach a few hundred because the modes are normalized in
the weighted operator norm. The identity-metric variant
(Fig.~\ref{fig:operator-gs}, Appendix~\ref{app:identity}) gives the
same qualitative picture on the Frobenius scale, with negative
eigenvalues reaching $\approx-0.85$. This is a direct numerical
demonstration of why the direct amplitude form \eqref{eq:C-direct}
may not be used without an eigenvalue check, whereas the projection
form \eqref{eq:proj}--\eqref{eq:blocks} and the squared form
\eqref{eq:C-safe} remain safe by construction. We also note that the
operator-space projection of the posterior covariance onto the
surviving modes compresses $A_1$ into a handful of scalar
coefficients (far cruder
than the vector-space bookkeeping, which retains the full mode-space
covariance $A_{\mathrm{proj}}$) so its reconstruction residual
must accompany any quantitative use.

\begin{figure*}[!tp]
\centering
\includegraphics[width=0.95\textwidth]{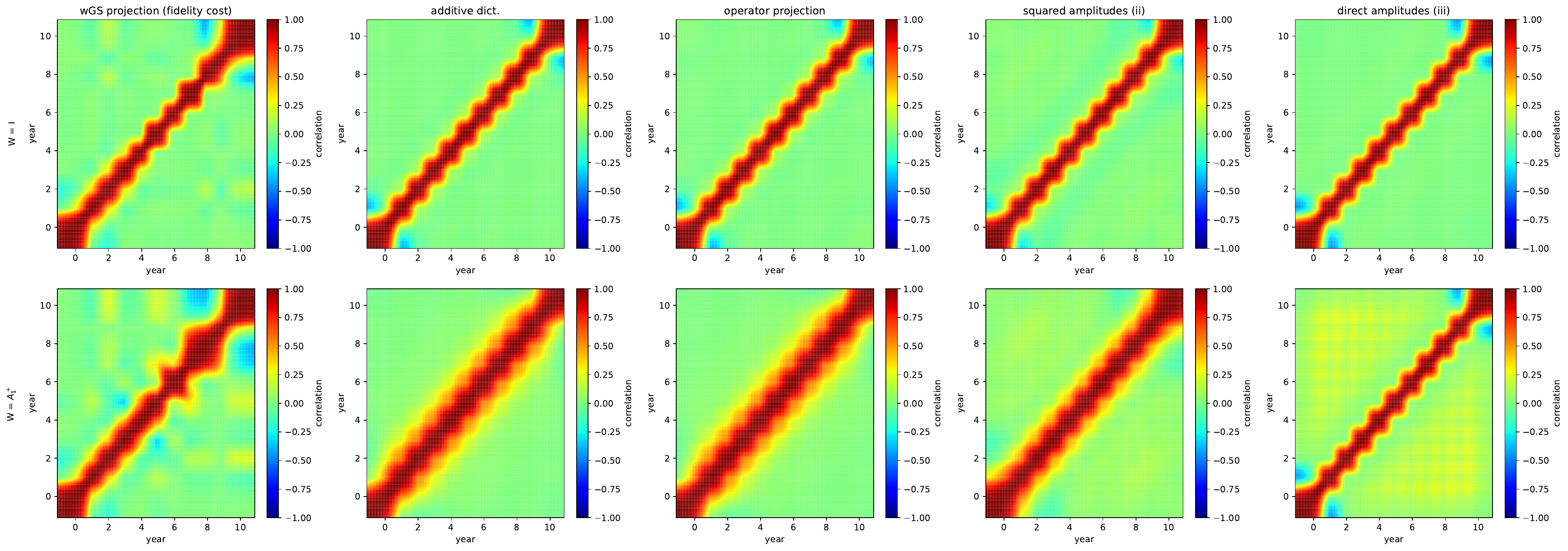}
\caption{Reconstructed correlation structure of the
AirPassengers FBET posterior on the prediction grid, for five
bookkeeping variants applied to the same posterior pair.
\textit{Columns, left to right:} the implemented weighted
Gram--Schmidt projection bookkeeping, the additive dictionary
\eqref{eq:naive}, the operator-space projection
(Sect.~\ref{sec:versionB}), the squared amplitude form (ii)
\eqref{eq:C-safe}, and the direct amplitude form (iii)
\eqref{eq:C-direct}. \textit{Top row:} declared baseline metric
$W=I$; \textit{bottom row:} posterior-precision metric
$W = A_1^{+}$, under which orthogonalization,
projections, and amplitude fits are all carried out in the
corresponding weighted (operator) norm. Under $W=I$ the projection
form retains the rich block structure while the two-amplitude
assemblies collapse most of the structure into a narrow diagonal
band. Under the posterior-precision metric the whitened fit target
is exactly the identity, so the amplitude assemblies attribute
almost everything to the nugget by construction, and the cross-scale
block of the projection bookkeeping vanishes identically (whitening
property, Sect.~\ref{sec:diagresults}).}
\label{fig:assemblies}
\end{figure*}

\paragraph{Alternative assemblies on the real posterior.}
Figure~\ref{fig:assemblies} compares the reconstructed
correlation structure of five bookkeeping variants (the
implemented projection bookkeeping, the additive dictionary
\eqref{eq:naive}, the operator-space projection of
Sect.~\ref{sec:versionB}, and the two amplitude assemblies of
Sect.~\ref{sec:assembly}) fitted to the same posterior
covariance, under both the declared baseline metric $W=I$ and the
posterior-precision metric $W = A_1^{+}$; in the
latter case orthogonalization, projections, and amplitude fits are
all carried out in the corresponding weighted (operator) norm. Under
the baseline metric the projection form retains the rich block
structure of the posterior, including anticorrelated off-diagonal
regions and the extrapolation-induced saturation at late times,
while the squared form (ii) and the direct form (iii), with only two
per-scale amplitudes and a nugget, reduce to a narrow positive band
around the diagonal with a weak periodic texture: most of the
posterior variance is absorbed by the nugget, and the cross-scale
information is, by construction of (ii), unavailable. Under the
posterior-precision metric the whitened fit target is exactly the
identity, so the amplitude assemblies degenerate to near-pure nugget
by construction, the additive dictionary broadens its attributed
band, and the projection variants remain informative but must be
read as an attribution of the whitened (not the raw) posterior
(cf.\ the whitening property above). Two numbers from this run make
the contrast concrete: the Frobenius norm of the wGS cross-scale
block drops from $\approx8.2\times10^{3}$ under $W=I$ to zero (exact)
under the posterior-precision metric, and the additive dictionary
attributes short- and long-scale components of norm
$\approx8.0\times10^{4}$ and $\approx7.7\times10^{4}$, each
exceeding the reconstructed total of $\approx4.6\times10^{4}$, i.e.\
the two components largely cancel, the numerical face of the
collinearity discussed in Sect.~\ref{sec:whygs} (under the
posterior-precision metric they inflate further, to
$\approx1.9\times10^{5}$). In the baseline run the
eigenvalue check of assembly (iii) passed without a nugget repair,
but that is a property of the fitted amplitudes, not a guarantee.

\begin{figure*}[!tp]
\centering
\includegraphics[width=0.8\textwidth]{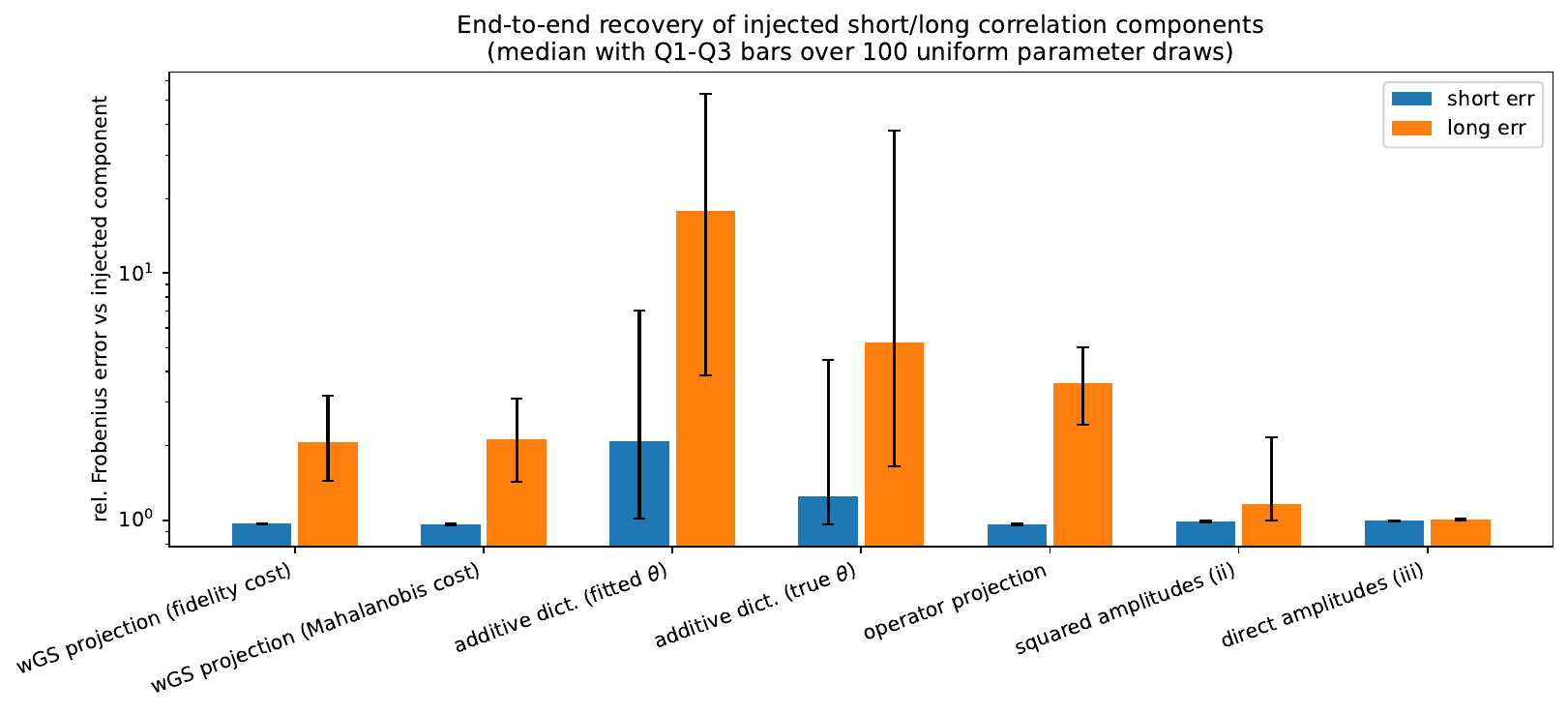}
\caption{End-to-end Monte Carlo recovery of injected short-
and long-scale correlation components for the AirPassengers test
case (median relative Frobenius
error with Q1--Q3 interquartile bars over 100 uniform draws of the
ground-truth hyperparameters; see Table~\ref{tab:simrecovery}). Every
method runs on the same simulated sample per trial, including both
hyperparameter-search cost conventions of the real pipeline, and all
outputs are evaluated on the same 200-node prediction grid.}
\label{fig:simrecovery}
\end{figure*}

\paragraph{End-to-end Monte Carlo recovery.}
Figure~\ref{fig:simrecovery} and Table~\ref{tab:simrecovery}
summarize the end-to-end simulation study: in each of 100 trials the
twenty ground-truth hyperparameters are drawn from uniform
distributions, a ground-truth correlation matrix
$C_{\mathrm{true}} = a_S \overline{K}_S + a_L \overline{K}_L + a_0 I$
is built on the monthly axis, one realization is simulated, the full
FBET stage is run, and every bookkeeping variant (including both
hyperparameter-search cost conventions of the real pipeline) recovers the short and long components, which are compared with the
injected ones on the common 200-node prediction grid. Three findings
stand out. First, every orthogonalized method has a
short-scale error close to
unity: the $n_g = 11$ evaluation grid, with node spacing of about one
year, cannot resolve sub-annual structure, so the injected seasonal
component is essentially invisible to the entire chain, a
grid-resolution limit, not a bookkeeping failure, and a concrete
argument for choosing $n_g$ against the shortest correlation length
of interest. Second, the additive dictionary with fitted
hyperparameters is unstable exactly as predicted: its long-scale
error has a median of $\approx18$ with an interquartile range
spanning from a few to above fifty, and even its short-scale error is
unstable (median $\approx2$, upper quartile $\approx7$, against
$\approx1$ for every other method), while
the same dictionary with the true hyperparameters performs
markedly better ($\approx5$) but retains a heavy upper tail (upper
quartile $\approx38$); the instability is
entirely due to the near-degenerate fitted basis, not to the
additive form itself. Third, the orthogonalized variants are stable
but conservative: the weighted Gram--Schmidt projection bookkeeping
reaches a long-scale error of
$\approx2.1$ under both cost conventions (statistically
indistinguishable), the direct amplitude form sits at unity
with a tight interquartile range, and the squared form near unity
with a broader upper quartile ($\approx2.2$), corresponding to
near-zero attributed
long-range amplitude. The choice between the additive-dictionary and
the orthogonalized
bookkeeping is thus a bias--variance trade-off made quantitative:
unbiased but wildly unstable attribution versus stable attribution
that systematically reassigns overlap.

\begin{table*}[!tp]
\caption{End-to-end Monte Carlo recovery for the AirPassengers
test case: median [Q1, Q3] of
the relative Frobenius error of the recovered short- and long-scale
correlation components over 100 uniform draws of the ground-truth
hyperparameters. ``Additive dictionary'' denotes the naive model
\eqref{eq:naive} of Sect.~\ref{sec:whygs} in its family
generalization (one amplitude per family member plus nugget, fitted
on the raw two-index kernels), with the kernel hyperparameters taken
either from the pipeline search (fitted $\theta$) or from the
simulation truth (true $\theta$).}
\label{tab:simrecovery}
\centering

\begin{tabular}{lcc}
\hline\hline
Method & Short err & Long err \\
\hline
wGS projection, fidelity cost \eqref{eq:projcost} & 0.97 [0.96, 0.97] & 2.07 [1.44, 3.20] \\
wGS projection, Mahalanobis cost & 0.96 [0.96, 0.97] & 2.12 [1.43, 3.10] \\
additive dictionary, fitted $\theta$ & 2.08 [1.01, 7.03] & 17.86 [3.85, 53.19] \\
additive dictionary, true $\theta$   & 1.25 [0.96, 4.45] & 5.23 [1.65, 37.66] \\
operator projection (Sect.~\ref{sec:versionB}) & 0.96 [0.96, 0.97] & 3.59 [2.43, 4.99] \\
squared amplitude form (ii)  & 0.99 [0.98, 0.99] & 1.16 [1.00, 2.16] \\
direct amplitude form (iii)  & 1.00 [0.99, 1.00] & 1.00 [1.00, 1.01] \\
\hline
\end{tabular}
\end{table*}

\subsection{Diagnostics for the luminosity-function case}
\label{sec:lfdisc}

\begin{figure*}[!tp]
\centering
\includegraphics[width=0.9\textwidth]{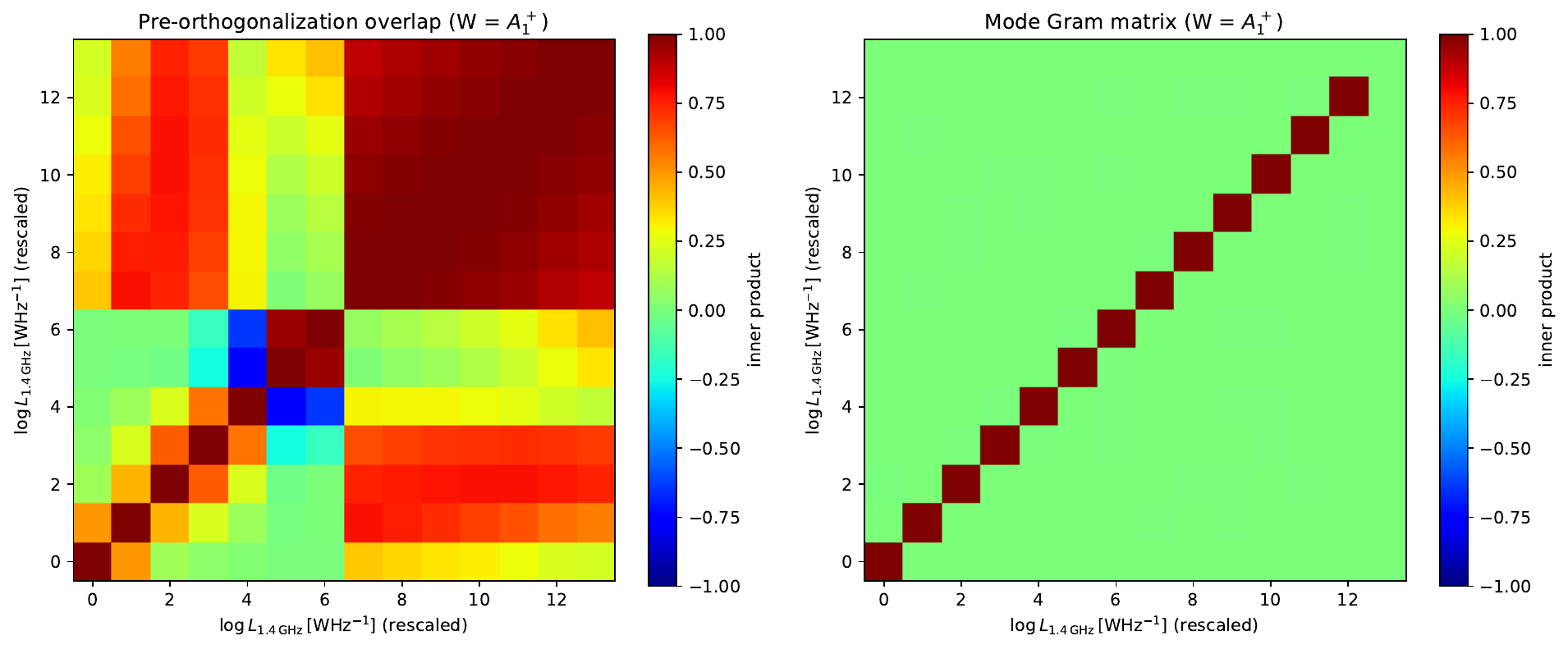}
\caption{Luminosity-function double-counting diagnostic under
the posterior-precision metric $W = A_1^{+}$:
pre-orthogonalization overlap of the fourteen raw basis functions
(\textit{left}; indices $0$--$6$ short family, $7$--$13$ long
family) and Gram matrix of the orthonormalized modes
(\textit{right}). The long-range family is internally
near-degenerate; the whitening property of
Sect.~\ref{sec:diagresults} applies verbatim. The identity-metric
variant is shown in Fig.~\ref{fig:lf-gram-overlap}
(Appendix~\ref{app:identity}).}
\label{fig:lf-gram-overlap-winv}
\end{figure*}

\begin{figure*}[!tp]
\centering
\includegraphics[width=0.95\textwidth]{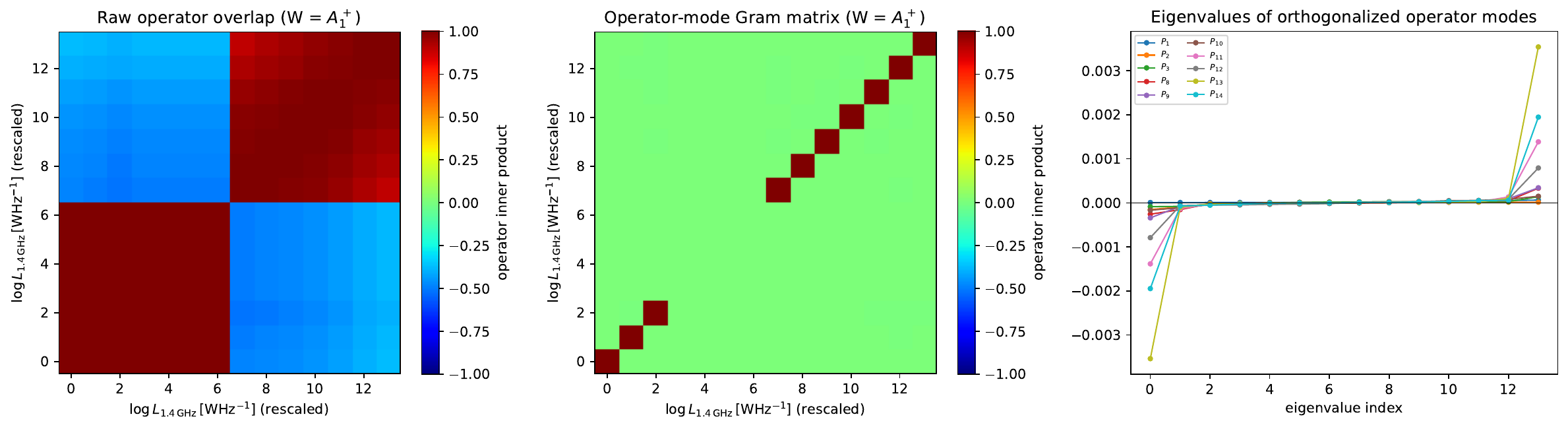}
\caption{Operator-valued (Version B) check for the
luminosity-function case under the posterior-precision metric
$W = A_1^{+}$, with the unwrapped short-range
distance. \textit{Left:} raw operator overlaps, the short-range
family is operator-identical, and the short--long cross overlaps are
strongly negative under this metric. \textit{Middle:} operator-mode
Gram matrix; empty diagonal slots are degenerate directions kept as
zero operators by the rank-aware recursion. \textit{Right:}
eigenvalue spectra of the surviving modes (magnitudes $\sim10^{-3}$
in the weighted normalization); all but the first are indefinite.
The identity-metric variant is shown in
Fig.~\ref{fig:lf-operator-gs} (Appendix~\ref{app:identity}).}
\label{fig:lf-operator-gs-winv}
\end{figure*}

\paragraph{Overlap structure and operator check.}
Figure~\ref{fig:lf-gram-overlap-winv} shows the pre-orthogonalization
overlap of the fourteen raw basis functions, under the
posterior-precision metric: the long-range family is
internally near-degenerate (mutual overlaps $\gtrsim 0.9$ over most
of the family), the anchored short-range bumps are comparatively well
separated (a banded structure with modest off-diagonal overlaps
and mild negative overlaps induced by the metric), and
the cross overlaps grow toward the large-$x$ end of the short family.
The Gram matrix of the surviving modes equals the identity, and the
whitening property of
Sect.~\ref{sec:diagresults} applies verbatim: the projected
mode-space covariance is exactly the identity on the kept modes, so
the cross-scale block vanishes by construction.
The identity-metric baseline
(Fig.~\ref{fig:lf-gram-overlap}, Appendix~\ref{app:identity})
shows the same degeneracy and cross-overlap pattern. The
operator-valued check (Fig.~\ref{fig:lf-operator-gs-winv}) sharpens the
picture: in the two-index realization, which depends on the
correlation lengths alone, the entire short-range family collapses to
operator overlaps of $\approx 1$, four of the fourteen operator
modes are degenerate (kept as zero slots by the rank-aware
recursion), and the surviving modes are indefinite (spectra of
magnitude $\sim10^{-4}$--$10^{-3}$ in the weighted normalization,
with only the
leading mode positive semidefinite), the same qualitative behavior
as for AirPassengers, in a more extreme regime.
A qualitatively new feature of this metric is that the
short--long cross overlaps, mildly positive under $W=I$
(Fig.~\ref{fig:lf-operator-gs}, Appendix~\ref{app:identity}, where
the surviving modes reach negative eigenvalues of $\approx-0.65$),
change sign and become strongly negative, under the precision
metric the two operator families are anti-aligned.

\begin{figure*}[!tp]
\centering
\includegraphics[width=0.85\textwidth]{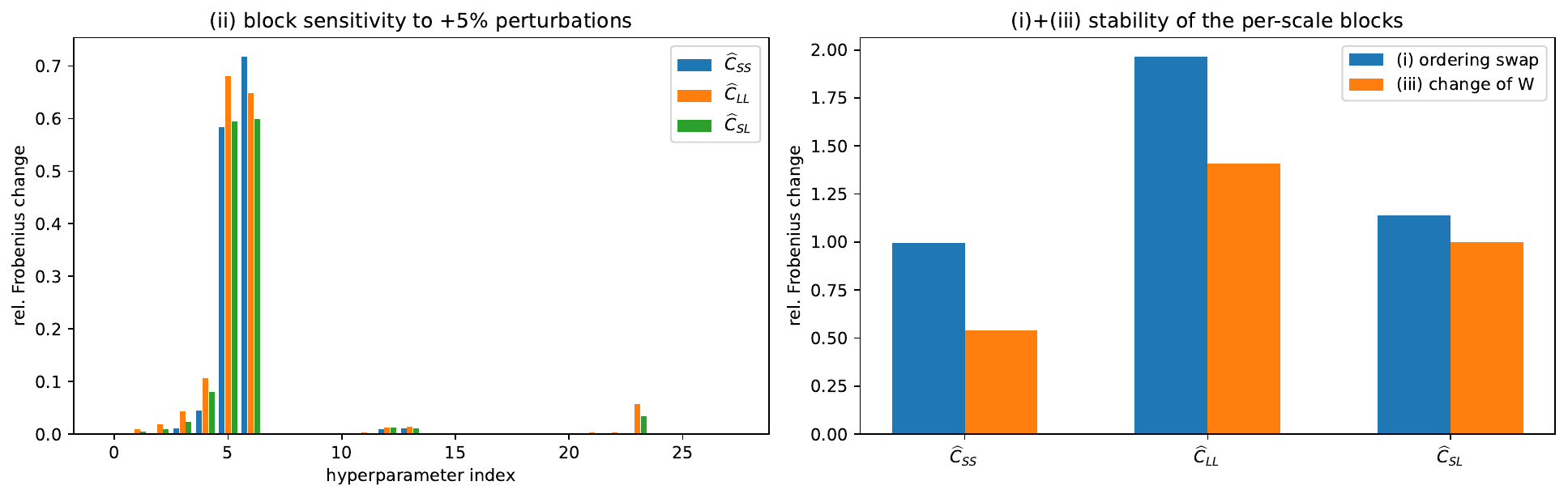}
\caption{Robustness of the luminosity-function per-scale
blocks. \textit{Left:} relative Frobenius change under a $+5\%$
perturbation of each of the twenty-eight basis-function
hyperparameters, the response is concentrated in a few short-family reference
positions, and most hyperparameters have
negligible influence. \textit{Right:} changes under a swap of the
orthogonalization ordering and under the change of the declared
metric from $W=I$ to the posterior precision
$A_1^{+}$; the long-range block dominates both.}
\label{fig:lf-robustness}
\end{figure*}

\paragraph{Robustness.}
The robustness scan (Fig.~\ref{fig:lf-robustness}, \textit{left})
concentrates the
sensitivity in a handful of short-family reference positions, with
relative block changes up to $\approx0.7$ for a $+5\%$ perturbation,
while most of the twenty-eight hyperparameters have negligible
influence.
The convention scan (Fig.~\ref{fig:lf-robustness},
\textit{right}) is again led by the long-range block: swapping the
orthogonalization ordering changes
$\widehat{C}_{\mathcal{L}\mathcal{L}}$ by a relative factor of
$\approx2$ (against $\approx1$ for the other blocks), and
replacing $W=I$ by the posterior-precision metric changes it by
$\approx1.4$. On this shared coordinate the long-range attribution
thus remains the most convention-dependent output, at a level
comparable to the AirPassengers case.

\begin{figure}[!tp]
\centering
\includegraphics[width=\columnwidth]{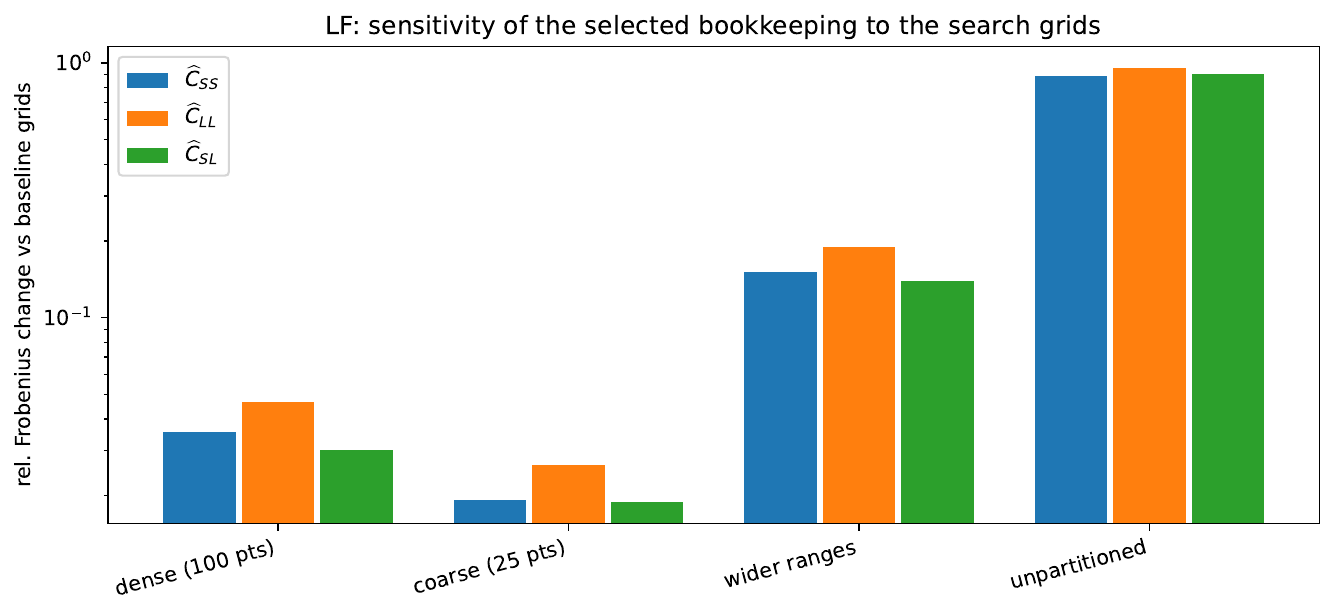}
\caption{Sensitivity of the selected luminosity-function
bookkeeping to the declared search grids (logarithmic scale), in the
same arrangement as Fig.~\ref{fig:grid-sensitivity}: relative
Frobenius change of the per-scale blocks against the baseline grids
for doubled density, halved density, widened ranges, and
unpartitioned anchor grids. The blocks are nearly insensitive to
density and range changes but change wholesale (by a factor of
$\approx25$ for $\widehat{C}_{\mathcal{L}\mathcal{L}}$) when the
anchor partition is removed.}
\label{fig:lf-grid-sensitivity}
\end{figure}

\paragraph{Grid sensitivity.}
The grid-sensitivity protocol
(Fig.~\ref{fig:lf-grid-sensitivity}; logarithmic scale) shows the
selected bookkeeping to be markedly more grid-stable than in the
AirPassengers case: doubled or halved grid density moves the
per-scale blocks by only a few per cent (relative Frobenius changes
of $\approx0.02$--$0.05$), widened ranges shift them by
$\approx0.13$--$0.19$, and removing the anchor partition changes
all three blocks by $\approx0.9$. The partition-of-ranges
convention thus remains the single most consequential ingredient of
the grid specification, but its influence is now of order unity,
comparable to the ordering and metric conventions of the robustness
scan.

\begin{figure*}[!tp]
\centering
\includegraphics[width=0.95\textwidth]{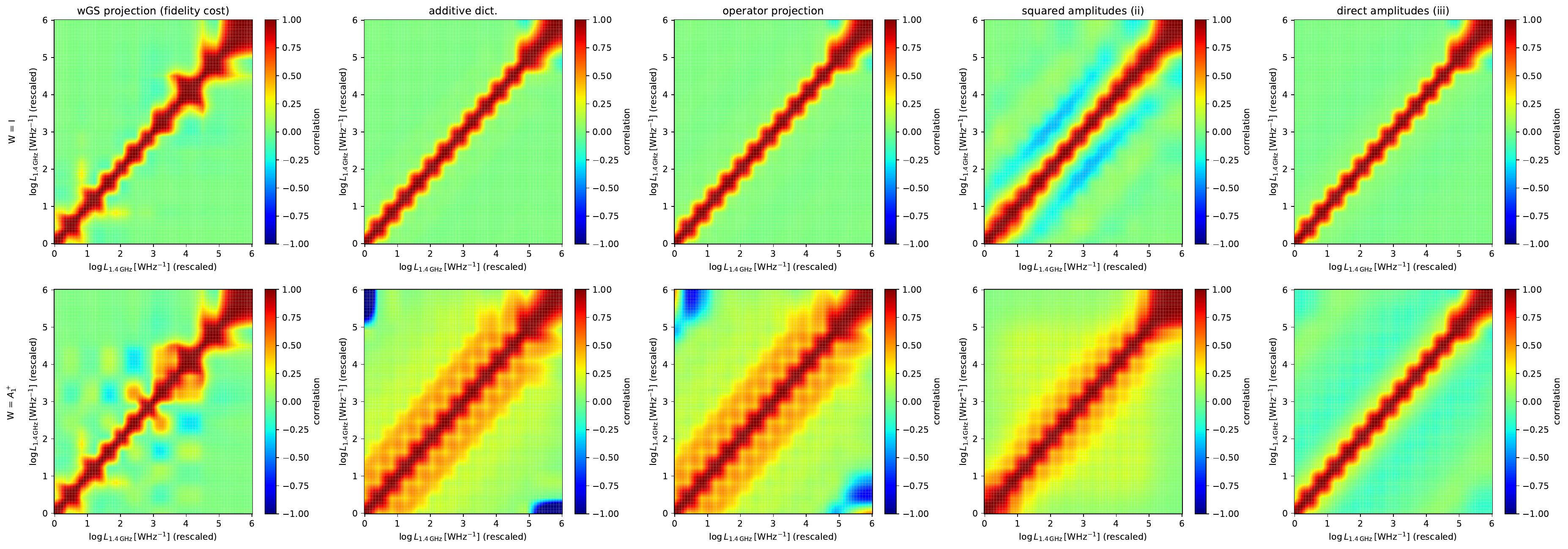}
\caption{Reconstructed correlation structure of the weighted
luminosity-function posterior, in the same arrangement as
Fig.~\ref{fig:assemblies}: \textit{columns} show the weighted
Gram--Schmidt projection bookkeeping, the additive dictionary, the
operator-space projection, the squared amplitude form (ii), and the
direct amplitude form (iii); \textit{rows} show the declared baseline
metric $W=I$ (top) and the posterior-precision metric
$W = A_1^{+}$ (bottom), under which all
orthogonalizations, projections, and amplitude fits are carried out
in the corresponding weighted norm.}
\label{fig:lf-assemblies}
\end{figure*}

\begin{figure*}[!tp]
\centering
\includegraphics[width=0.8\textwidth]{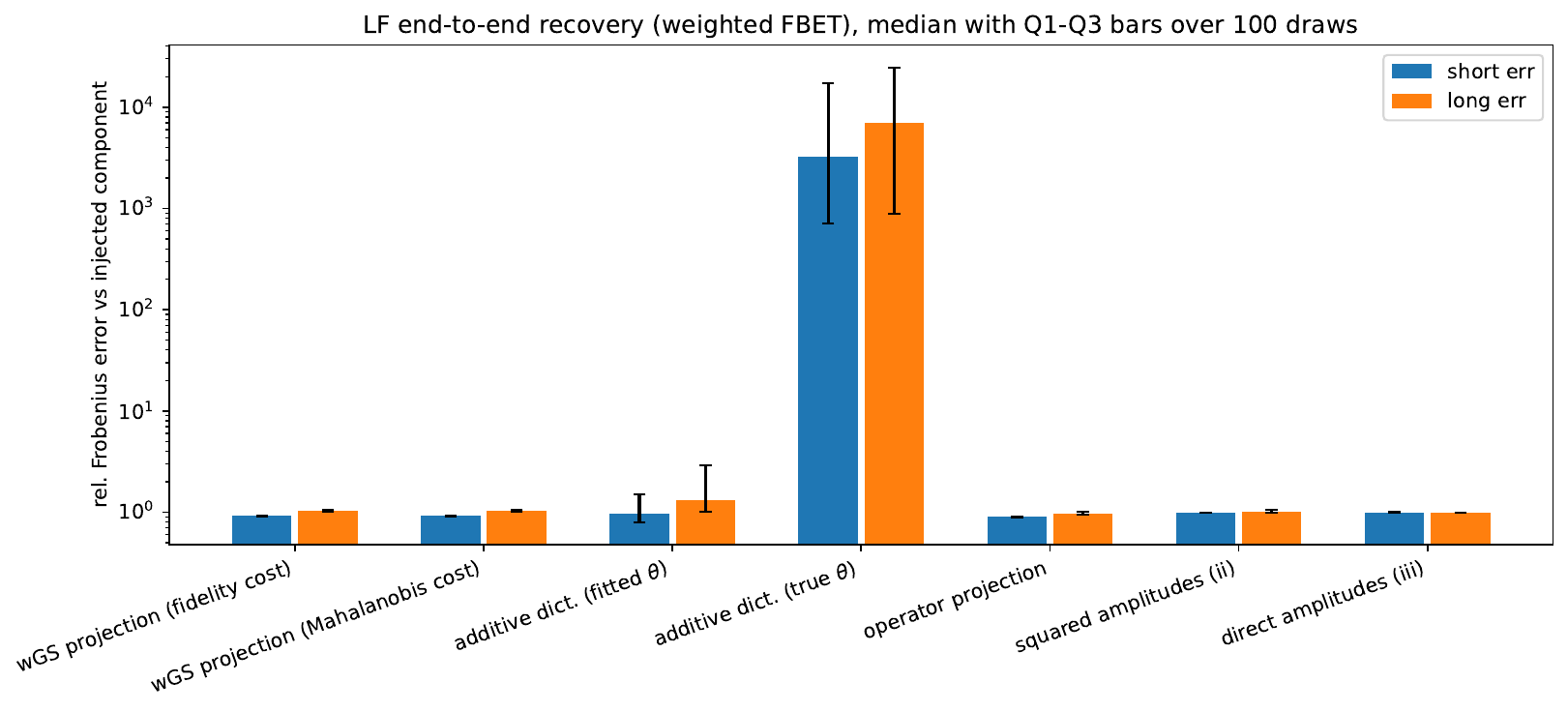}
\caption{End-to-end Monte Carlo recovery for the
luminosity-function case (weighted FBET stage; median relative
Frobenius errors with Q1--Q3 bars over 100 uniform draws;
\emph{logarithmic} scale). On this non-cyclic abscissa the additive
decomposition with the true generating hyperparameters is
catastrophically unidentifiable (three to four orders of magnitude
above unity), while its fitted-hyperparameter variant and every
orthogonalized variant remain stable near unity.}
\label{fig:lf-simrecovery}
\end{figure*}

\paragraph{Assemblies and end-to-end recovery.}
The assembly comparison on the real weighted posterior
(Fig.~\ref{fig:lf-assemblies}; five methods $\times$ two metrics, as
in Fig.~\ref{fig:assemblies}) mirrors the AirPassengers result under
the baseline metric: the
projection bookkeeping retains the posterior structure, the squared
form develops anticorrelated flanks along the diagonal, and the
direct form reduces to a narrow positive band. Under the
posterior-precision metric the additive dictionary and the operator
projection broaden their attributed correlation bands markedly,
while the amplitude assemblies stay near the diagonal: the
whitened fit target is the identity, and the projection variants must
be read as attribution of the whitened posterior. The end-to-end Monte
Carlo (Fig.~\ref{fig:lf-simrecovery}; note the logarithmic scale)
delivers the strongest message of the entire study: on a non-cyclic
abscissa, where both scales act through the \emph{same} lag variable
and are separated only by the functional form of the kernels, the
additive decomposition with the \emph{true} generating
hyperparameters becomes catastrophically unidentifiable: its recovery
errors sit three to four orders of magnitude above unity, with
interquartile ranges spanning more than a decade. The same
decomposition with \emph{fitted} hyperparameters stays near unity:
the search selects kernel parameters that avoid the collinear
configuration, whereas the true generating kernels, sharing one lag
axis, produce a strongly collinear design; the collinearity is
structural, a property of the generating configuration rather than of
the estimation. All
orthogonalized variants (both weighted Gram--Schmidt projection
costs, the operator projection, and both amplitude assemblies) remain
stable with errors near unity. Together with
Table~\ref{tab:simrecovery}, this completes the bias--variance
picture: on a cyclic coordinate the additive model with true
hyperparameters is a competitive upper bound, on a shared coordinate
it fails outright, while the orthogonalized bookkeeping behaves
consistently in both regimes.

\subsection{Assessment}

\paragraph{Advantages.} Within its stated scope, the construction
offers: an interpretable separation of wrapped-seasonal and
long-range chronological effects, with orthogonality holding by
construction under the declared metric; avoidance of naive covariance
addition and of the double counting it entails; compatibility with
information-geometric weighting frameworks built on
marginalized weights \citep{imbrisak2025},
since the same weighted-inner-product machinery is used throughout;
and a covariance parametrization whose per-scale amplitudes support
uncertainty propagation with auditable scale attribution. In the
projection form \eqref{eq:proj}--\eqref{eq:blocks} the cross-scale
covariance block is reported rather than assumed to vanish, turning a
common silent assumption into an inspectable output, and positive
semidefiniteness holds by congruence without an eigenvalue check.

\paragraph{Limitations.} The limitations are real and should be
weighed explicitly. The basis choice matters: $\kS$ and $\kL$ are two
specific basis functions among many defensible ones, and conclusions about
``the seasonal mode'' are conditional on the basis function family. The
choice of $W$ matters: orthogonality is metric-relative, and a
different statistically motivated $W$ yields different modes; nothing
in the construction selects $W$ automatically. Positive definiteness
must be checked: the direct assembly \eqref{eq:C-direct} can fail
positive semidefiniteness after orthogonalization, and even the safe
form \eqref{eq:C-safe} requires verification once nuggets, taperings,
or truncations are applied. The orthogonalization depends on ordering
(the declared ordering convention) and must be re-run whenever $W$ is updated.
The implemented realization is finite-rank: the bookkeeping is exact
only within the span of the basis function family, structure orthogonal to
that span is invisible to it, and the reconstruction residuals of
Section~\ref{sec:assembly} must accompany any per-scale attribution.
The hyperparameter search of Section~\ref{sec:hyperopt} is a heuristic
with no optimality guarantee, and several of its conventions (the
projection-fidelity cost, the $\chi^2$ normalization inside the
channel weights, the truncated weighting-prior support) are declared
rather than derived. Finally, the test cases are illustrative, not
definitive: AirPassengers is a single, short, well-behaved series
chosen for interpretability, and the luminosity-function case, while
adding genuine channel structure, is likewise a single dataset; no
claim is made that behavior observed on either generalizes.

\paragraph{Relation to existing practice.} Additive kernel models are
standard in Gaussian-process regression \citep{rasmussen2006}, where
sums of kernels are legitimate covariance models by construction. The
present concern is different: when per-component \emph{amplitudes are
themselves the objects of inference and uncertainty attribution} (as in evaluation frameworks descended from generalized-least-squares nuclear data
evaluation \citep{schnabel2018,schnabel2021}) overlap between
components corrupts the attribution even when the summed covariance is
admissible. The bookkeeping step proposed here is aimed at that
attribution problem, not at replacing additive kernel modeling in
settings where attribution is not required.

\paragraph{Covariance parametrization (design-level
proposal).}
As a proposal for the next iteration of this framework (explicitly \emph{not} part of the implemented protocol) the
structured operator \eqref{eq:C-safe} (plus nugget) would provide a
low-dimensional, interpretable parametrization of $A_0$ or $B$ in
\eqref{eq:fbet}, with the per-scale amplitudes promoted to genuine
hyperparameters of the evaluation, selected within the FBET cycle in
direct analogy to the marginalized channel weights of \eqref{eq:Bw}.
This is where the bookkeeping would influence the inference itself
rather than only report on it: the implemented posterior projection
decomposes an already-computed posterior after the fact, whereas the
proposed parametrization would let the declared correlation scales
shape the prior and the update. Because the modes are orthonormal
under the working metric, the amplitude-estimation problem is
expected to be far better conditioned than under the naive dictionary
\eqref{eq:naive}; the Monte Carlo study of Sect.~\ref{sec:diagresults}
supplies the quantitative motivation, with amplitude fits on the raw
additive model ranging from unstable (with fitted hyperparameters
on the cyclic coordinate) to catastrophically
unidentifiable (with the true generating hyperparameters on the
shared coordinate; Table~\ref{tab:simrecovery},
Fig.~\ref{fig:lf-simrecovery}), while the orthonormalized variants
remain stable. The implementation
realizes the posterior bookkeeping only.

\section{Conclusion}
\label{sec:conclusion}

We have described a structured correlation bookkeeping construction
intended for weighted FBET-type uncertainty quantification. The
construction separates short-range wrapped seasonal correlations,
defined through a circular lag on the month coordinate, from
long-range chronological correlations, defined through the absolute
time lag, and treats the corresponding kernels as candidate basis functions
rather than as independent covariance components. A weighted
Gram--Schmidt procedure (in a vectorized form, an operator-valued
kernel form, or the finite-rank single-index realization used by the
implementation) converts the basis functions into modes that
are orthonormal under an explicitly declared inner product, after
which per-scale amplitudes may be assigned, or an existing FBET-type
posterior mean and covariance may be projected into per-scale and
cross-scale blocks and propagated. The naive additive model is retained only as
a labeled pre-orthogonal dictionary, never as the final construction.
Conceptually, the approach transfers the marginalized
weighting philosophy of \citet{imbrisak2025} from
dataset balancing to correlation-scale bookkeeping: covariance
structure is connected to a declared weighting geometry rather than
introduced as an ad hoc additive term. The AirPassengers and luminosity-function datasets provide small,
interpretable testbeds for the protocol and implementation
documented in Sections~\ref{sec:prototype}--\ref{sec:results}; the
full predeclared diagnostic suite is executed and reported in
Sects.~\ref{sec:results} and~\ref{sec:discussion}.

\section*{Data availability}
The AirPassengers series is publicly available as Series~G of
\citet{box2015} and is distributed, for example, with the R
statistical environment. The luminosity-function measurements are
from the VLA-COSMOS 3\,GHz Large Project \citep{novak2018}.


\appendix

\section{Identity-metric variants of the diagnostic figures}
\label{app:identity}

For completeness, this appendix collects the identity-metric
($W=I$) variants of the overlap and operator diagnostics whose
posterior-precision versions appear in the main text. The comparison
shows that the overlap structure (and hence the double-counting
risk) is a property of the basis-function family design rather
than of the declared metric.

\begin{figure*}[!tp]
\centering
\includegraphics[width=0.85\textwidth]{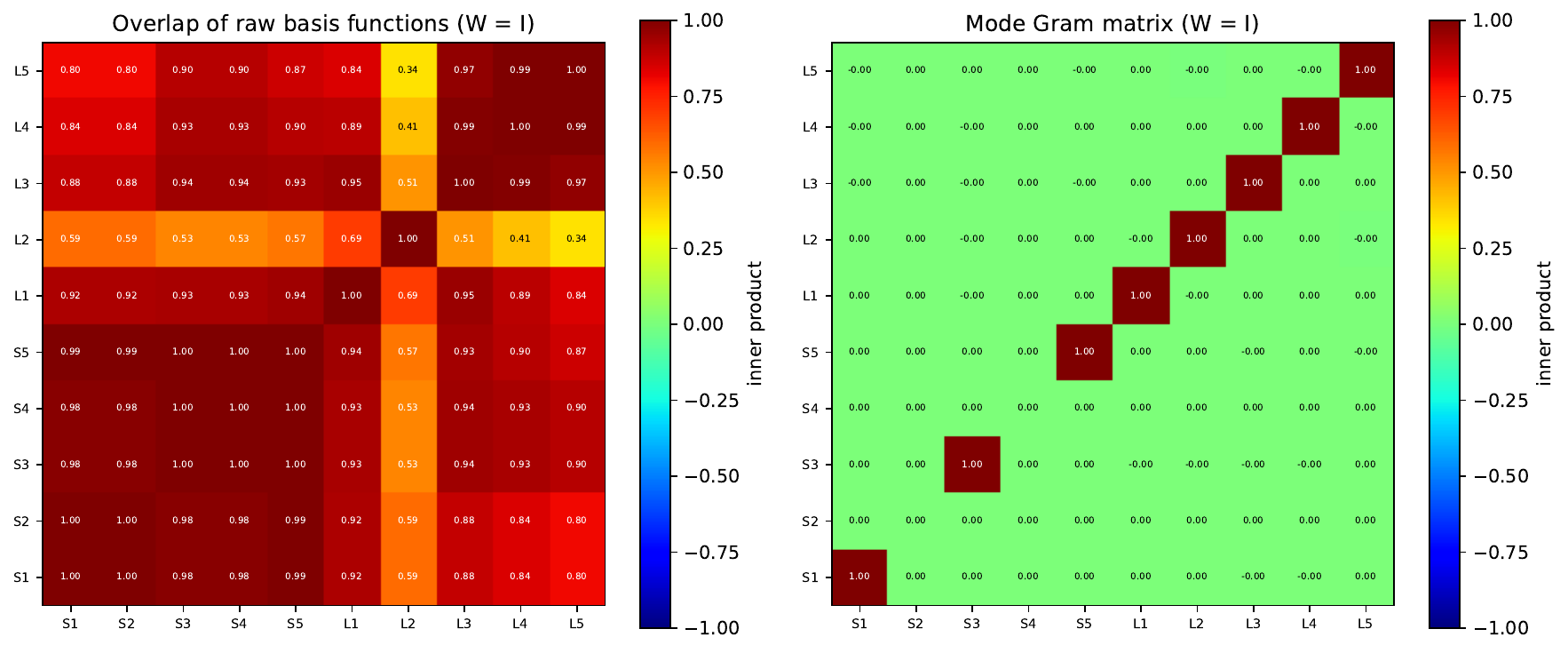}
\caption{Same as Fig.~\ref{fig:gram-overlap-winv}
(AirPassengers double-counting diagnostic), under the declared
baseline metric $W=I$. All pairwise overlaps lie between $0.34$ and
$1.00$; the overlap pattern is nearly identical to the
posterior-precision case, so the double-counting risk is a property
of the family design rather than of the metric.}
\label{fig:gram-overlap}
\end{figure*}

\begin{figure*}[!tp]
\centering
\includegraphics[width=0.95\textwidth]{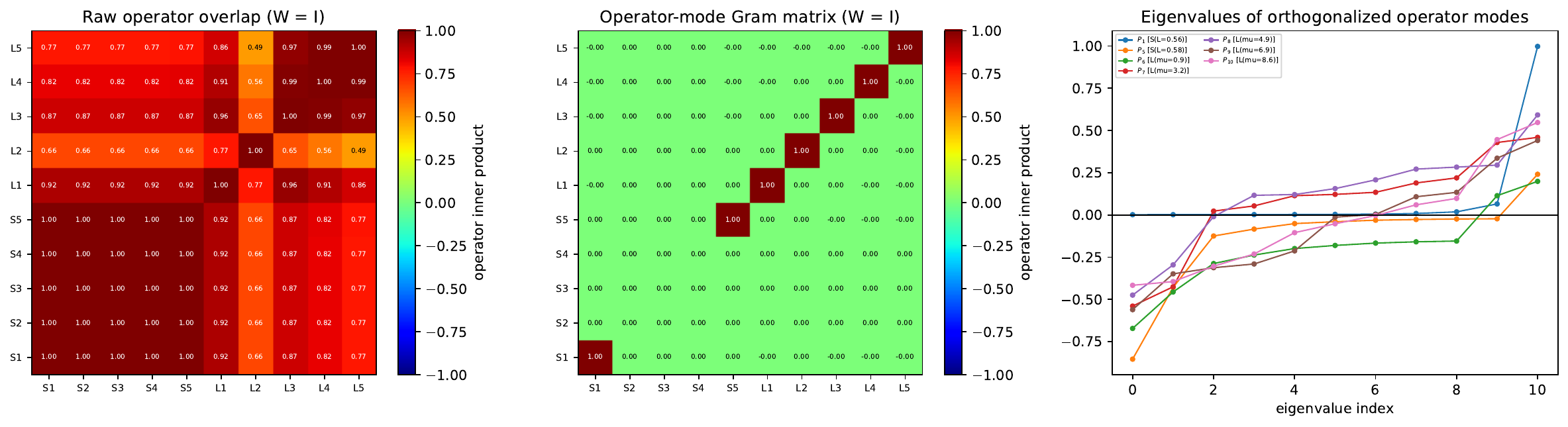}
\caption{Same as Fig.~\ref{fig:operator-gs-winv}
(AirPassengers operator-valued check), under the declared baseline
metric $W=I$. Seven of the ten operator modes survive the rank-aware
orthogonalization; only the first surviving mode is positive
semidefinite, with negative eigenvalues of the subsequent modes
reaching $\approx-0.85$, the direct numerical motivation for the
eigenvalue caveat on the direct amplitude form
\eqref{eq:C-direct}.}
\label{fig:operator-gs}
\end{figure*}

\begin{figure*}[!tp]
\centering
\includegraphics[width=0.9\textwidth]{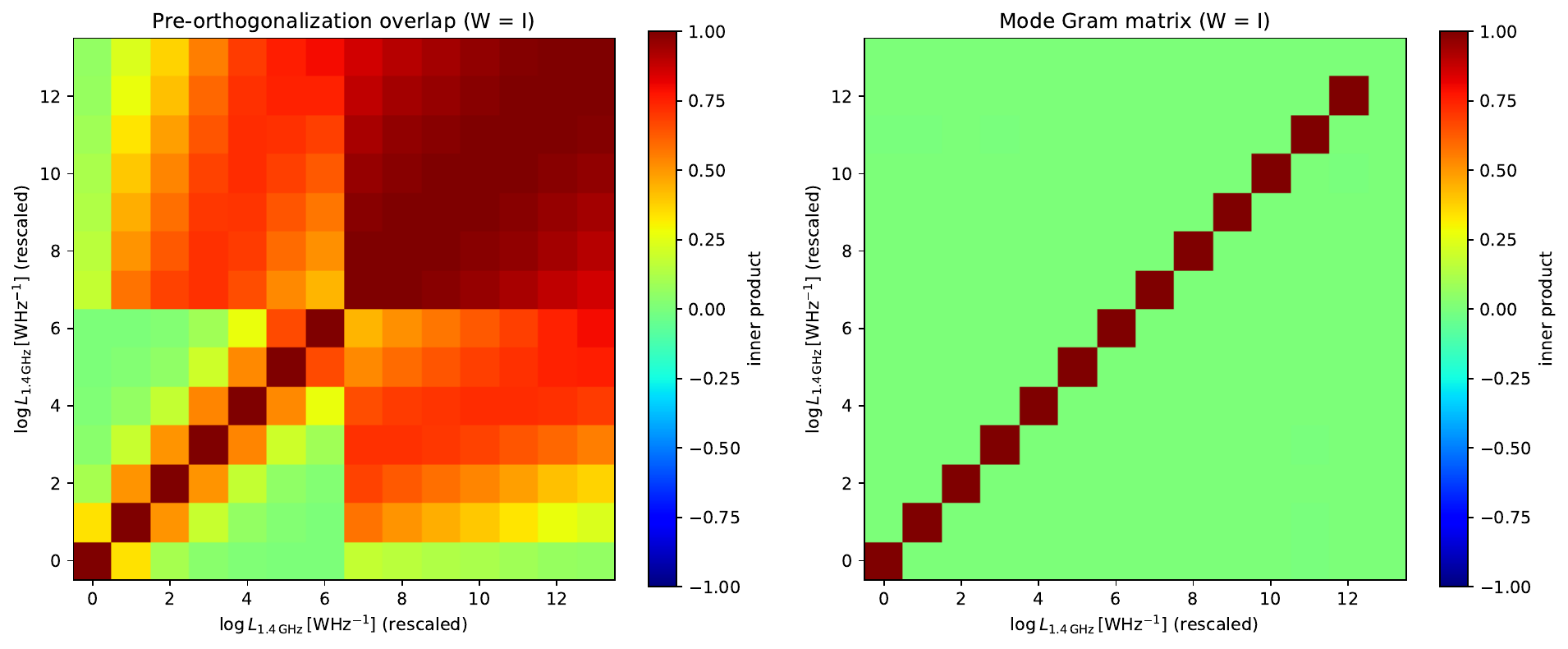}
\caption{Same as Fig.~\ref{fig:lf-gram-overlap-winv}
(luminosity-function double-counting diagnostic), under the declared
baseline metric $W=I$. The long-family degeneracy and the
cross-overlap pattern match the posterior-precision case; the short
family lacks the mild negative overlaps induced by the precision
metric.}
\label{fig:lf-gram-overlap}
\end{figure*}

\begin{figure*}[!tp]
\centering
\includegraphics[width=0.95\textwidth]{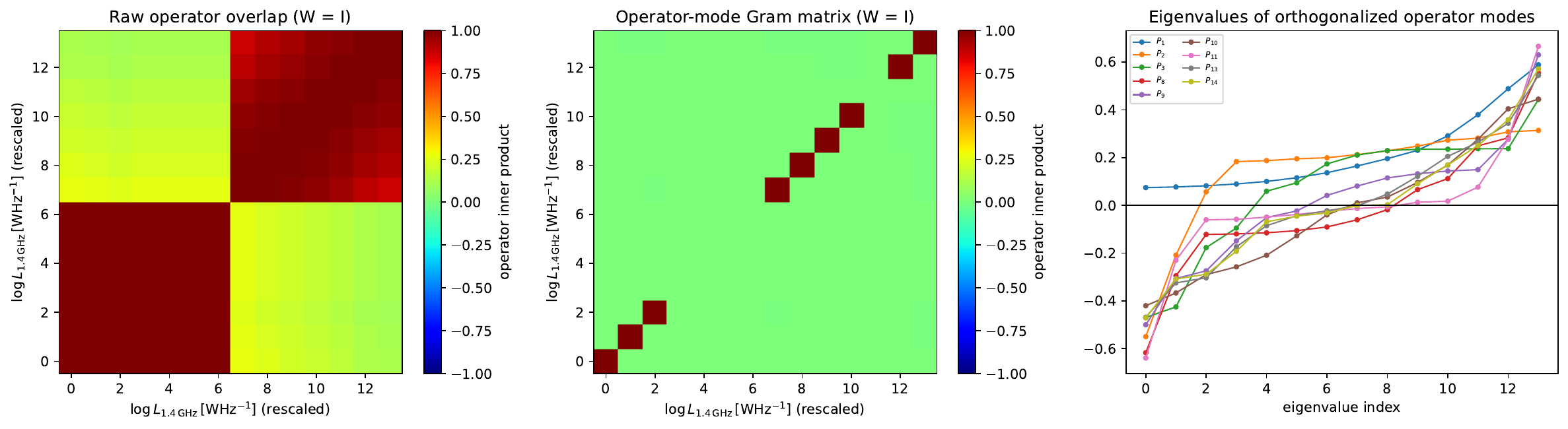}
\caption{Same as Fig.~\ref{fig:lf-operator-gs-winv}
(luminosity-function operator-valued check), under the declared
baseline metric $W=I$. The short--long cross overlaps are mildly
positive (they change sign under the precision metric), five of the
fourteen operator modes are degenerate zero slots, and the surviving
modes are indefinite with negative eigenvalues reaching
$\approx-0.64$.}
\label{fig:lf-operator-gs}
\end{figure*}


\begin{thebibliography}{99}

\bibitem[Beran(1994)]{beran1994}
Beran, J. 1994, Statistics for Long-Memory Processes, Monographs on
Statistics and Applied Probability, Vol.~61 (New York: Chapman \& Hall)

\bibitem[Box et al.(2015)]{box2015}
Box, G.~E.~P., Jenkins, G.~M., Reinsel, G.~C., \& Ljung, G.~M. 2015,
Time Series Analysis: Forecasting and Control, 5th edn.
(Hoboken: Wiley), source of the AirPassengers dataset (Series~G);
earlier editions: Holden-Day, 1970, 1976

\bibitem[Cressie(1993)]{cressie1993}
Cressie, N.~A.~C. 1993, Statistics for Spatial Data, revised edn.
(New York: Wiley)

\bibitem[Hobson et al.(2002)]{hobson2002}
Hobson, M.~P., Bridle, S.~L., \& Lahav, O. 2002, MNRAS, 335, 377

\bibitem[Hosking(1981)]{hosking1981}
Hosking, J.~R.~M. 1981, Biometrika, 68, 165

\bibitem[Imbri\v{s}ak et al.(2025)]{imbrisak2025}
Imbri\v{s}ak, M., Lovell, A.~E., \& Mumpower, M.~R. 2025,
arXiv:2508.19468

\bibitem[Imbri\v{s}ak \& Tisani\'c(2026)]{imbrisak2026}
Imbri\v{s}ak, M. \& Tisani\'c, K. 2026, arXiv:2604.10782

\bibitem[Leeb et al.(2008)]{leeb2008}
Leeb, H., Neudecker, D., \& Srdinko, T. 2008, Nuclear Data Sheets,
109, 2762

\bibitem[Novak et al.(2018)]{novak2018}
Novak, M., Smol\v{c}i\'c, V., Schinnerer, E., et al. 2018, A\&A, 614,
A47

\bibitem[Rasmussen \& Williams(2006)]{rasmussen2006}
Rasmussen, C.~E. \& Williams, C.~K.~I. 2006, Gaussian Processes for
Machine Learning (Cambridge, MA: MIT Press)

\bibitem[Schnabel(2018)]{schnabel2018}
Schnabel, G. 2018, arXiv:1803.00960

\bibitem[Schnabel et al.(2021)]{schnabel2021}
Schnabel, G., Capote, R., Koning, A.~J., \& Brown, D. 2021,
arXiv:2110.10322

\end{thebibliography}
\end{document}